\documentclass[12pt]{article}
\pdfoutput=1
\usepackage{graphicx}
\usepackage{jheppub}  
\usepackage[usenames,dvipsnames,table]{xcolor}
\usepackage{amsmath}
\usepackage{amsthm}
\usepackage{amsfonts}
\usepackage{braket}
\usepackage{bbold}
\usepackage{mathrsfs}  
\usepackage[mathscr]{eucal}
\usepackage{tikz}
\usepackage{enumerate}
\usepackage{stmaryrd}

\def\sect#1{Section \ref{#1}}
\def\fig#1{Fig.\,\ref{#1}}
\def\req#1{(\ref{#1})}

\definecolor{Acol}{rgb}{0.8, 0.2, 0.4}
\definecolor{Bcol}{rgb}{0.3, 0.7, 0.4}
\definecolor{Ccol}{rgb}{0.2, 0.3, 0.8}
\definecolor{Dcol}{rgb}{0.5, 0.5, 0.5}
\definecolor{ABcol}{rgb}{0.9, 0.6, 0.2}
\definecolor{BCcol}{rgb}{0.2, 0.7, 0.8}
\newtheorem{conjecture}{Conjecture}[section]

\newcommand{\Tr}{\textrm{Tr}}

\numberwithin{equation}{section}

\def\coop{cooperative }
\def\area#1{{\rm Area}\left[#1\right]}  
\def\m#1{{\mathfrak{m}_{#1}}}  
\def\mm#1#2{{\mathfrak{m}_{#1}^{#2}}}  
\def\extr#1{{\mathfrak{E}_{#1}}}  
\def\v#1{v_{#1}} 
\def\pv{{\tilde v}} 
\def\fto{{\shortrightarrow}}
\def\half{\frac{1}{2}}
\def\veps{\varepsilon}
\def\ph{\varphi}
\def\ep{\varphi}  

\def\la{2}
\def\lb{3}
\def\lc{1}
\def\quarter{{\frac{1}{4}}}

\def\locfol{local foliation}
\def\strand{thread bundle}
\def\strands{thread bundles}
\def\strnd{bundle}
\def\strnds{bundles}

\subheader{}
\title{\boldmath Bulk locality and cooperative flows}

\author{Veronika E.~Hubeny}
\affiliation{Center for Quantum Mathematics and Physics (QMAP)\\
Department of Physics, University of California, Davis, CA 95616 USA}

\emailAdd{veronika@physics.ucdavis.edu}

\abstract{
We use the `bit thread' formulation of holographic entanglement entropy to highlight the distinction between the universally-valid strong subadditivity and the more restrictive relation called monogamy of mutual information (MMI), known to hold for  geometrical states (i.e.\ states of holographic theories with gravitational duals describing a classical bulk geometry).  In particular, we provide a novel proof of MMI, using bit threads directly.  To this end, we present an explicit geometrical construction of {\it cooperative flows} which we build out of disjoint {\it \strands}.  We conjecture that our method applies in a wide class of configurations, including ones with non-trivial topology, causal structure, and time dependence.  The explicit nature of the construction reveals that MMI is more deeply rooted in bulk locality than is the case for strong subadditivity. 
}

\begin{document} 

\maketitle
\flushbottom
\newpage 

\section{Introduction}
\label{s:intro}

Entanglement, a quintessentially quantum characteristic, is expected to play a key role in holography, not only as a powerful tool in describing a given physical system, but also as a fundamental building block in the duality itself.  Understanding precisely how this works has been one of the outstanding goals in the holography program over the last few years.  The basic quantity of interest is entanglement entropy, which provides a particularly natural measure of entanglement, and from which other useful quantities are built.  Universal relations between these quantities allow us to gain further insights, and in the holographic context may even diagnose whether a given quantum field theory state has a holographic dual corresponding to a classical geometry.  The way in which the bulk geometry implements these relations can then provide crucial insights to unraveling the emergence of spacetime in holography.  

In the following, we will focus on one particular relation known as  monogamy of mutual information (MMI).  In order to explain its significance, we first review its construction in the broader context of other relations.\footnote{
In the interest of being reasonably self-contained, in the next few paragraphs we give a very brief review of these relations.  This is well-known material; for a nice review of the relations and their proofs in the holographic context (for time reflection symmetric configurations), see e.g.\ \cite{Headrick:2013zda}. 
}
The entanglement entropy (EE) of a system $A$ can be thought of as a measure of mixedness of its reduced density matrix $\rho_A$ obtained by tracing out the complement ${\bar{A}}$ in a specified bi-partitioning of the system
${\cal H} = {\cal H}_A \otimes {\cal H}_{\bar{A}}$.  Explicitly, EE is defined as the von Neumann entropy
\begin{equation}
S(A) = - \Tr  \, \rho_A  \, \log \rho_A \ . 
\label{e:EE}
\end{equation}	
Refining the partition into several subsystems $A$, $B$, etc., we can consider general constraints on the respective entanglement entropies.\footnote{
It is known that even in the restricted context of holography \cite{Bao:2015bfa}, there are in fact infinitely many such relations if we allow arbitrary number of partitions, though the complete set has hitherto remained elusive.
}
For example, among particularly useful relations are sub-additivity (SA),  strong sub-additivity (SSA) and its cousin (related by purification) weak monotonicity (WM), and the above-mentioned monogamy of mutual information (MMI), which can respectively be expressed as:
\begin{align}
\label{e:sumSA}
& \rm{(SA)}:  & S(A) + S(B) &\ge S(AB) \\
\label{e:sumSSA}
& \rm{(SSA)}: & S(AB) + S(BC)&\ge  S(B) + S(ABC) \\
\label{e:sumWM}
& \rm{(WM)}: & S(AB) + S(BC) &\ge   S(A) + S(C) \\
\label{e:sumMMI}
& \rm{(MMI)}: & S(AB) + S(BC) + S(CA) &\ge S(A) + S(B) + S(C) + S(ABC)
\end{align}
where  we use the shorthand $AB \equiv A \cup B$, etc. 
 While these relations look superficially similar, there is an important difference between the first three and the last one:  SA, SSA, and WM are all valid for any quantum system in any state and for any partition.  MMI, on the other hand, is a stronger relation which does NOT hold in general, but as initially proved in \cite{Hayden:2011ag}, it does hold for holographic states.  
 
To digest the significance of this statement, it is useful to rewrite the above relations in a slightly more suggestive way.
One can think of the triangle inequality type relation of SA \req{e:sumSA} as the statement of positivity of the mutual information
\begin{equation}
I(A:B) \equiv S(A) + S(B) - S(AB) \ ,
\label{e:MIdef}
\end{equation}	
which characterizes the amount of correlations (both classical and quantum) between the systems $A$ and $B$.  
One can similarly view the convexity property of SSA \req{e:sumSSA} as the statement of  positivity of the conditional mutual information
\begin{equation}
I(A:C|B) \equiv I(A:BC) - I(A:B) \ ,
\label{e:CMIdef}
\end{equation}	
which says that the amount of correlation is monotonic under inclusion.  These two statements, that the amount of correlations cannot be negative and that enlarging the system cannot decrease the amount of correlations with another system, are intuitively clear.\footnote{
Though perhaps surprisingly so; see \cite{Witten:2018zva} for an exposition of some of the subtleties.
}

The statement of MMI \req{e:sumMMI}, on the other hand, translates to the statement of negativity of the tripartite information\footnote{
One could argue that `negative tripartite information' (also known as interaction information) is a more natural quantity to consider, since its positivity appears in a variety of QI contexts on similar footing as other positive quantities.  However, to avoid confusion, we will presently stick to the usual convention of using the standard $I_3$.
}
\begin{equation}
I_3(A:B:C) \equiv I(A:B)+I(A:C)-I(A:BC) \ ,
\label{e:tripIdef}
\end{equation}	
which can therefore be thought of as measuring the non-extensivity of mutual information.  If the correlations between the three systems are purely quantum, then they must obey the property known as monogamy of entanglement (implying that quantum entanglement between $A$ and $B$ cannot be shared by $C$ etc.), so the entanglement contributing to $I(A:B)$ cannot comprise any part of contribution to $I(A:C)$ and vice-versa.  That in turn means that $I(A:BC)$ must be at least as large as the individual contributions, $I(A:B)+I(A:C)$, implying the MMI relation \req{e:sumMMI} (and motivating its name).  If, on the other hand, the correlations are classical, then they {\it can} be redundantly shared, allowing for the possibility of positive tripartite information.\footnote{
In fact even the $n\ge 4$ party GHZ state $\frac{1}{\sqrt{n}} \,  \left( |00 \ldots 0 \rangle +  |11 \ldots 1 \rangle \right) $, which in some regards is highly quantum but whose correlations are maximally redundant, violates MMI. 
}
  In the holographic context, then, the fact that MMI holds (i.e.\ that $I_3\le0$) is rather suggestive of sufficient quantumness of the correlations between subsystems.\footnote{
This has already been suggested by \cite{Hayden:2011ag} and  is further supported by the recent exploration of \cite{Takayanagi:2018zqx} using the relative entropy of entanglement as a better measure of quantumness of correlations.
}
This emphasizes the important point that while we tend to think of the bulk as behaving classically, this is merely an effective description: the corresponding CFT state manifests quantum features even with respect to the observables naturally associated with the bulk variables.

One of the advantages of the bulk description is that it often conveniently geometrizes highly non-trivial statements about the system, allowing them to be proved with ease, at least in a certain context.\footnote{
In everything that follows, we will focus on the leading order in the large-$N$, large-$\lambda$ limit of the CFT, which corresponds to a classical geometry in the bulk.  (However we briefly comment on  quantum and stringy correction in \sect{s:disc}.)
}
The inequalities \req{e:sumSA}--\req{e:sumMMI} provide an excellent example of this intriguing fact.
According to the Ryu-Takayanagi (RT) prescription  \cite{Ryu:2006ef,Ryu:2006bv} pertaining to static situations, the entanglement entropy of a given spatial region $A$ in the boundary CFT\footnote{
Although in our gauge theory context, the Hilbert space factorization for such a spatial partitioning strictly-speaking does not apply, this does not affect any of the results discussed below,
as can be justified using the more rigorous framework of Tomita-Takesaki theory based on algebra of observables;  for a recent review see e.g.\ \cite{Witten:2018zxz}.
} is computed by quarter of the proper area (in Planck units) of a  minimal surface\footnote{
The surface is by definition of  co-dimension-two in the full bulk spacetime, which in the static context can be more usefully viewed as spatially co-dimension-one within a constant time slice.
} 
$\m{A}$  which is homologous to $A$ (and hence anchored on the entangling surface $\partial A$):
\begin{equation}
S(A) = \frac{1}{4 \, G_N} \, \area{\m{A}} \ .
\label{e:RT}
\end{equation}	
In  arbitrary time-dependent situations, this relation naturally generalizes to the Hubeny-Rangamani-Takayanagi (HRT) prescription \cite{Hubeny:2007xt}, wherein the minimal surface $\m{A}$ at constant time gets simply replaced by a spacetime co-dimension-two extremal surface $\extr{A}$ homologous to $A$ 
(and in case of multiple such candidates, being the one with the smallest area).
However, in what follows we will primarily focus on the static case.

Remarkably, the proof of SSA in the static case, first obtained in \cite{Headrick:2007km} and later generalized in \cite{Hayden:2011ag}, follows almost immediately from \req{e:RT}.  As we briefly review in \sect{ss:setup} in a simple setup, one can obtain \req{e:sumSSA} by comparing the respective bulk minimal surface areas and using the fact that a minimal surface by definition cannot have larger area than any other surface in the same homology class.  Interestingly, the argument for MMI is virtually identical \cite{Hayden:2011ag} -- indeed, as indicated below, in this context the proof of MMI actually reduces to that of SSA, despite the fact that these are logically distinct relations with different physical meanings.
This curious statement likewise holds in the general time-dependent setting \cite{Wall:2012uf}.  In other words, the RT and HRT prescriptions do not discern the fundamental distinction between SSA and MMI.   Hence their virtue, that the bulk repackaging simplifies the dual CFT description, turns into somewhat of a detriment, by obscuring a crucial aspect of these relations.\footnote{
There are several other  features, such as the bulk location of the CFT information, for which these prescriptions likewise seem more puzzling than suggestive.}

To elucidate this important distinction, then, we need to go beyond the RT (or HRT) prescription.  Of course the structural difference between SSA and MMI would become more apparent if we include quantum corrections, but going beyond the classical limit does not address the essence of the puzzle directly.  Fortuitously, however, we in fact {\it can} see a distinction already at the leading order in large-$N$ description, as explained below.  This is manifested by utilizing a different geometrical prescription for holographic entanglement entropy, namely the ``bit thread" formulation of Freedman and Headrick \cite{Freedman:2016zud}.
  This prescription, reviewed in \sect{s:proof}, captures the quantum information theoretic meaning of the relevant constructs more naturally.\footnote{
Although, as argued in \cite{Freedman:2016zud} using the Max Flow-Min Cut theorem and shown more explicitly in \cite{Headrick:2017ucz}, the bit thread prescription for holographic entanglement entropy is mathematically equivalent to the RT prescription, it nevertheless turns out to be much better tailored to the present considerations.
}
  Here entanglement entropy is given by the maximum flux of a bulk `flow' (defined as a divergence-free vector field of bounded norm), whose integral curves are the bit threads.  As shown in \cite{Freedman:2016zud},  
  the statements of SA and SSA can be recast into statements of positivity of the number of threads obeying certain restrictions -- and since this number manifestly cannot be negative, these two relations are then even more directly evident in the bit thread formulation than via RT.

However, as already emphasized in  \cite{Freedman:2016zud}, the MMI relation does {\it not} lend itself to such an easy argument; in fact, \cite{Freedman:2016zud} has not been able to find a proof directly in the flow language without recourse to the equivalence with RT.  
The bit thread formulation, then, provides a useful window of opportunity: a framework wherein the distinction between SSA and MMI should become more directly apparent, which we can use to elucidate the role of the geometrization and ultimately the emergence of bulk spacetime.
More specifically, \cite{Freedman:2016zud} demonstrates that just using the basic building blocks of the flows (and its crucial properties such as `nesting') does not guarantee MMI: some other non-trivial property of flows is needed.  In this paper, we set out to elucidate this property.  We will find that it is deeply rooted in bulk locality.

In particular, we prove MMI directly using the language of bit threads.\footnote{
The authors of \cite{MattMMI} (whom we thank for sharing an advance version of their draft) likewise have a proof of MMI,  inspired by previously known theorems regarding multicommodity flows, using tools of the theory of convex optimization (in particular the Lagrangian duality) along similar lines as \cite{Headrick:2017ucz}, to abstractly argue for the existence of `\coop flows'.  Our argument is complementary to that line of reasoning: it is explicitly constructive and directly geometrical.
}
The structure of the presentation is as follows.
Our argument starts with the observation\footnote{
We thank Matt Headrick for sharing this observation, which inspired the present work (as well as the term `\coop flows'). 
}
 (explained in \sect{s:proof}),
that under the assumption of the existence of certain `\coop flows', MMI would follow easily.  
The crux of the argument then is to demonstrate the existence of such \coop flows.  We achieve this by an explicitly constructive method.  In \sect{s:simple} we explain the basic idea in a simple context, and in \sect{s:generaliz} we generalize this construction to arbitrary spatial partitions of CFT states describing arbitrary asymptotically AdS geometries.  
We then revisit the relation to bulk locality and discuss further implications and open questions in \sect{s:disc}.

\section{Bit thread proof of MMI -- overview}
\label{s:proof}

To approximate a partitioning of the Hilbert space ${\cal H} = {\cal H}_A \otimes {\cal H}_B \otimes {\cal H}_C \otimes {\cal H}_D$ (with $D \equiv \overline{ABC}$ being the purifier)  in the holographic context, we typically let $A$, $B$,  $C$, and $D$ correspond to spatial regions partitioning the space on which the theory lives.\footnote{
\label{fn:wormholes}
For the CFT living on a single connected spacetime, this restriction corresponds to taking a pure state, so the bulk admits no eternal black hole horizons.  However, it is easy to generalize to the case with black holes (or even more generally `multi-boundary wormholes') by letting $D$ (as well as $A$, $B$, and $C$) include regions in multiple disconnected boundary spacetimes.
}
Recall that the general form of MMI, written in a manifestly $ABC$-cyclically symmetric form, is
\begin{equation}
S(AB) + S(BC) + S(CA) \ge S(A) + S(B) + S(C) + S(ABC) \ ,
\label{e:MMI}
\end{equation}	
where each term gives the entanglement entropy of the corresponding boundary subregion.

In this section, we outline the basic argument for demonstrating \req{e:MMI} in the holographic context using the bit thread formulation.  
  Following \cite{Freedman:2016zud}, we can describe the entanglement entropy in terms of bulk flows.  A {\it flow} $v$ is defined as divergenceless vector field of bounded norm, 
\begin{equation}
\nabla \cdot v =0 \ , \quad |v| \le 1 \ ,
\label{e:flowdef}
\end{equation}	
whose integral curves are dubbed `bit threads'. 
The entanglement entropy 
of a given spatial region $X$ on the boundary is computed by maximizing the flux through $X$ over all bulk flows, 
\begin{equation}
S(X) = \max_v \int_X v \ ,
\label{e:EEflowdef}
\end{equation}	
where we're using the shorthand 
$\int_X v := \int_X \sqrt{h} \, n_\mu \, v^\mu = \int_\m{X} \sqrt{h} \, n_\mu \, v^\mu $ (with $n_\mu$ being the unit normal, and $h$ the determinant of the induced metric, on the integration surface which by homology and Stokes' theorem can be taken at $\m{X}$).
Any flow which maximizes the flux through a region $X$ is called 
a {\it maximizer flow} for the region $X$ and denoted by\footnote{
 In the interest of compactness of notation, here we deviate from  \cite{Freedman:2016zud} (which uses  the notation $v(X)$ for the maximizer flow).  To label multiple generic flows which are not necessarily maximizing for any region, we will use a numerical subscript, e.g.\ $v_1, v_2$, etc.\ (not to be confused with a letter subscript which labels a region and signifies maximizer flow through that region).
 } $\v{X}$, so the entanglement entropy can be characterized more simply as the flux of a maximizer flow,
\begin{equation}
S(X) = \int_X \v{X} \ge \int_X v \ ,
\label{e:EEflowrel}
\end{equation}	
where the inequality by definition holds for any flow $v$.
Note that a maximizer flow $\v{X}$ for a given region $X$ is highly non-unique: the restriction \req{e:flowdef} only fully fixes it  at the minimal surface  $\m{X}$ (through which it must pass perpendicularly and with unit norm).  For later purposes it is also useful to note that switching the direction of given a maximizer flow $\v{X}$ for region $X$ automatically generates a maximizer flow $\v{{\overline X}}=-\v{X}$ for its complement ${\overline X}$.

Applying \req{e:EEflowrel} to the LHS of the MMI inequality \req{e:MMI},
we have
\begin{equation}
\begin{split}
S(AB) + S(BC) + S(CA)
&= \int_{AB} \v{AB} + \int_{BC} \v{BC} + \int_{CA} \v{CA} \\
&\ge
\int_{AB} v_{\lc} + \int_{BC} v_{\la} + \int_{CA} v_{\lb} \\
&= \int_{A} (v_{\lc} + v_{\lb}) + \int_{B} (v_{\lc}+v_{\la}) + \int_{C} (v_{\la} + v_{\lb}) \ ,
\end{split}
\label{e:LHSbound}
\end{equation}
where the inequality holds for arbitrary flows $v_{\lc}$, $v_{\la}$, and $v_{\lb}$.
To relate this to the RHS of \req{e:MMI}, we want to choose these flows $v_i$ (with $i = 1,2,3$) to be given by 
\begin{equation}
\begin{split}
v_{\lc} = \pv_\lc &\equiv  \half \left(   \v{A} + \v{B} - \v{C} + \v{ABC} \right) \\
v_{\la} = \pv_\la &\equiv  \half \left( -  \v{A} + \v{B} + \v{C} + \v{ABC} \right) \\
v_{\lb} = \pv_\lb &\equiv  \half \left(  \v{A} - \v{B} + \v{C} + \v{ABC} \right) 
\end{split}
\label{e:v123def}
\end{equation}
where $\v{A}$, $\v{B}$, $\v{C}$, and $\v{ABC}$ are maximizer flows for the corresponding regions.
It is easy to see that in such a case the RHS of \req{e:LHSbound} would reduce to
\begin{equation}
\begin{split}
& \int_{A} ( \pv_{\lc}+\pv_{\lb}) + \int_{B} ( \pv_{\lc}+\pv_{\la}) + \int_{C} (\pv_{\la} + \pv_{\lb}) \\
&=
 \int_{A} (\v{A}+\v{ABC}) + \int_{B}(\v{B}+\v{ABC}) + \int_{C} (\v{C}+\v{ABC})) \\
&= S(A) + S(B) + S(C) + S(ABC) \ ,
\end{split}
\label{e:RHS}
\end{equation}
thereby proving the MMI relation.

The crux of the argument then boils down to showing that the objects $\pv_i$ defined by \req{e:v123def} are indeed flows, i.e.\ that they satisfy \req{e:flowdef} and hence we're allowed to equate them to the LHS terms in \req{e:v123def}.  While divergencelessness of $\pv_i$'s is guaranteed by linearity, the norm bound need not a-priori remain satisfied:  By triangle inequality we're only guaranteed that the average $\half \, (v_\alpha+v_\beta)$ of any two flows $ v_\alpha$ and $v_\beta$ is itself a flow, whereas the expressions involving 4 flows in \req{e:v123def} can exceed unit norm
(and in fact one can easily construct maximizer flows $\v{A}$, $\v{B}$, $\v{C}$, and $\v{ABC}$, for which each of $|\pv_\lc|$, $|\pv_\la|$, and $|\pv_\lb|$  becomes 2 somewhere).

However, not all maximizer flows $\v{A}$, $\v{B}$, $\v{C}$, and $\v{ABC}$ have this undesirable property.  Given any  specified regions $A$, $B$, and $C$, we can find 
 maximizer flows $\v{A}$, $\v{B}$, $\v{C}$, and $\v{ABC}$ which guarantee that $|\pv_\lc|\le 1$, $|\pv_\la|\le 1$, and $|\pv_\lb|\le 1$ simultaneously throughout the bulk.
 Below, we outline an explicit construction 
(in fact two natural ones) of maximizer flows
  which satisfy this requirement,
  thus supplying the missing step in the proof of MMI.  
Our construction utilizes the observation \cite{Headrick:2017ucz} that a foliation of a bulk region by minimal surfaces induces a maximally-collimated flow; we demonstrate that suitable foliations always exist which `comb' the flows $\v{A}$, $\v{B}$, $\v{C}$, and $\v{ABC}$ in such a way as to `cooperate' in the requisite fashion.

\section{Basics of \coop flow construction}
\label{s:simple}

We start with a simple class of examples to illustrate the basic idea, and subsequently generalize our arguments to more complicated situations in \sect{s:generaliz}.
In particular, we first fix convenient dimensionality and state, but consider a generic partition within a given topology class.  

\subsection{Setup}
\label{ss:setup}

\begin{figure}[htbp]
\begin{center}
\includegraphics[width=2.1in]{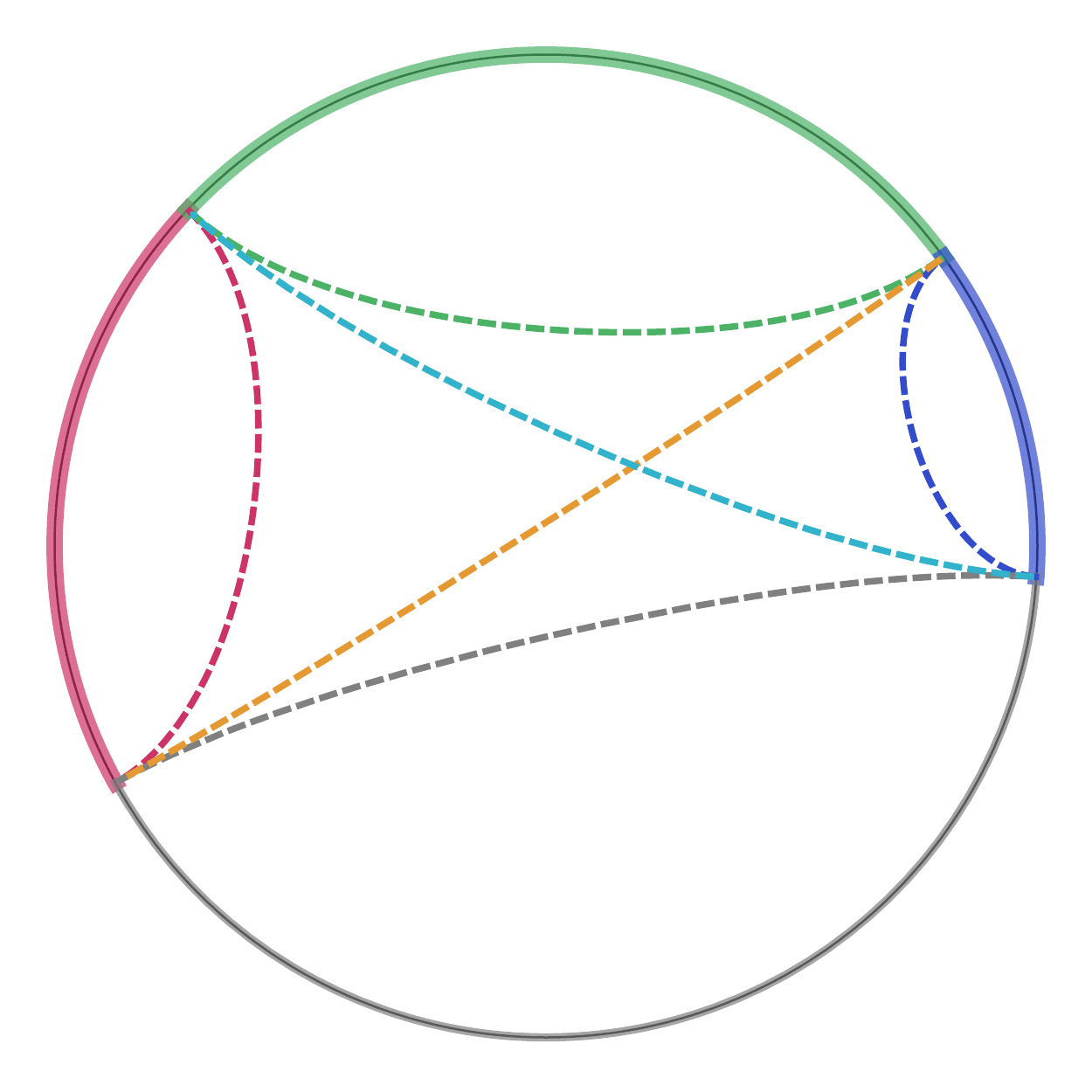}
\put(-136,74) {\makebox(20,20) {\footnotesize {\color{Acol}{$\m{A}$}}}}
\put(-78,102) {\makebox(20,20) {\footnotesize {\color{Bcol}{$\m{B}$}}}}
\put(-44,80) {\makebox(20,20) {\footnotesize {\color{Ccol}{$\m{C}$}}}}
\put(-70,48) {\makebox(20,20) {\footnotesize {\color{Dcol}{$\m{D}$}}}}
\put(-104,68) {\makebox(20,20) {\footnotesize {\color{ABcol}{$\m{AB}$}}}}
\put(-105,83) {\makebox(20,20) {\footnotesize {\color{BCcol}{$\m{BC}$}}}}
\put(-162,80) {\makebox(20,20) {{\color{Acol}{$A$}}}}
\put(-90,143) {\makebox(20,20) {{\color{Bcol}{$B$}}}}
\put(-11,90) {\makebox(20,20) {{\color{Ccol}{$C$}}}}
\put(-50,-2) {\makebox(20,20) {{\color{Dcol}{$D$}}}}

\caption{An example of partitioning of the boundary space into regions (thick curves) $A$, $B$, $C$, and the complement $\overline{ABC}\equiv D$, as labeled.  The corresponding minimal surfaces (dashed curves) for these 4 regions, along with those for $AB$ and $BC$ are also indicated.   (In this and subsequent plots, we plot the Poincare disk in global coordinates described by  constant $t$ slice of $ds^2 = \frac{1}{\cos^2 \rho} \left[-dt^2+d\rho^2 + \sin^2 \rho \, d\ph^2\right]$ with $\rho = \pi/2$ corresponding to the boundary.)}
\label{f:simple}
\end{center}
\end{figure}
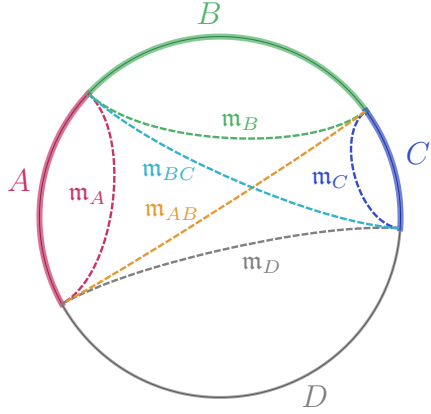

Let us consider pure AdS$_3$ geometry, with all regions of interest localized at a fixed time $t=0$, so that we can WLOG restrict attention to bulk spatial slice, i.e.\ the Poincare disk.\footnote{
We will only use the metric details of AdS$_3$ for generating the explicit plots, 
but otherwise our arguments will be robust to deforming the geometry.  Moreover, as long as we're restricting to static context, our results will not depend on assuming any specific field equations, energy conditions, etc..
}
We will consider the boundary to live on $S^1$ space and take the partitions $A$, $B$, and $C$ to be simple adjoining intervals.  We will  denote the complementary region $\overline{ABC}\equiv D$ and assume the total state to be pure, i.e.\ $S(ABCD)=0$.
A representative of a generic such configuration is sketched in \fig{f:simple}, along with the corresponding minimal surfaces relevant for the terms appearing in  MMI \req{e:MMI}.
Note that
$S(ABC) = \quarter \, \area{\m{D}}$ and
\begin{equation}
S(AC) = 
\quarter \, \min\{\area{\m{A}}+\area{\m{C}} , \area{\m{B}}+\area{\m{D}} \} \  .
\label{e:SAC}
\end{equation}	
We can now readily confirm that the RT-based geometric proof of MMI is then manifestly identical\footnote{
Note that our present argument based on cancelation of surfaces hinges on the fact that our `minimal surfaces' are one-dimensional; in higher dimensions one has to recourse to the more refined argument of \cite{Hayden:2011ag}.
} to the SSA proof of \cite{Headrick:2007km}: for either option in  the RHS of \req{e:SAC}, $S(AC)$ is canceled by two of the terms on the RHS of \req{e:MMI}, with the remaining terms equivalent to either SSA \req{e:sumSSA},
 or its purification WM \req{e:sumWM}.
In the present configuration these two relations follow from cutting and rejoining $\m{AB}$ and $\m{BC}$ so as to implement $AC$ homology, and observing that each surface necessarily has greater area than the globally minimal surface in its homology class, namely $\m{B}$ and $\m{D}=\m{ABC}$ in the case of \req{e:sumSSA}, or  $\m{A}$ and $\m{C}$ in the case of \req{e:sumWM}.
In the following construction we will for definiteness\footnote{
As we further explain in footnote \ref{fn:IACgen}, this is not a limitation since an analogous method will work for the opposite sign of the inequality in \req{e:BDneck} as well.  Moreover, MMI is preserved under purification and permutations, so the choice \req{e:BDneck} merely fixes a labeling convention.
 \label{fn:IAC}} pick the case
\begin{equation}
\area{\m{B}}+\area{\m{D}}  \ge \area{\m{A}}+\area{\m{C}} 
\label{e:BDneck}
\end{equation}	
so that
$S(AC)=S(A)+S(C)$. 
(Although this implies $I(A:C)=0$, we will develop our proof of MMI without the reliance on this fact.)

Converting the above statements to the bit thread description \cite{Freedman:2016zud,Headrick:2017ucz}, the entanglement entropy of each region is  given by the flux of the corresponding maximizer flow.  Recall that any maximizer flow for  the region $A$ is then guaranteed to be maximally packed (i.e.\ $|\v{A}|=1$) on $\m{A}$, and similarly for the regions $B$, $C$, and $ABC$, which are maximally packed on the corresponding minimal surfaces.
 Elsewhere these individual flows admit some floppiness, and at the AdS boundary they are maximally floppy.  Hence there is a large amount of freedom in picking the  maximizer flows $\v{A}$, $\v{B}$, $\v{C}$, and $\v{ABC}$.  However, as explained above,  a generic choice would not necessarily satisfy the flow conditions \req{e:flowdef} for the $\pv_i$'s defined in \req{e:v123def}; our task is to find an explicit construction which would guarantee that $|\pv_\lc|\le 1$, $|\pv_\la|\le 1$, and $|\pv_\lb|\le 1$ simultaneously throughout the bulk.

To build up some intuition, we first take a short detour to illustrate why constructing  flows is not trivial, and in particular requires some  global considerations.  We will however start in \sect{ss:detour} with an observation that the set up indicated in \sect{ss:setup} actually contains a hidden simplification, which we could use to prove the desired result in this particular case.   We will then explain in \sect{ss:plan} why this method is not amenable to an easy generalization, and motivate the general construction, which will simultaneously reveal the connection to bulk locality.  To skip to the construction itself, the reader may proceed directly to \sect{ss:constr}.

\subsection{Cooperative flows for uncorrelated regions}
\label{ss:detour}

In encountering the requisite expressions \req{e:v123def}, one might be tempted to try  reducing each line to only two terms by canceling the other two against each other.  
Unfortunately, we can not set, say, $\v{A}=\v{B}$, since  by subadditivity, such flow across $\m{A}$ and $\m{B}$ would exceed the common bottleneck $\m{AB}$,\footnote{
This statement holds whenever subadditivity is not saturated (i.e.\ when $I(A:B)>0$, since then $S(A)+S(B)>S(AB)$), and in the present case, the mutual information $I(A:B)$ in fact diverges.  (Geometrically this is associated with the UV divergence in the areas of $\m{A}$ and $\m{B}$ at the common entangling surface $A\cap B$, which is not present for $\m{AB}$.)
}  and similarly for the other adjoining regions.  The only simultaneous flows we are allowed a-priori (without additional constraints), are the ones corresponding to nested regions, i.e.\ $ABC$ and (at most) one of $A$, $B$, or $C$.  This  however  ensures the norm bound for only one of the $\pv_i$'s.  To deal with the other two $\pv_i$'s, we need further cancellations.  

In fact, one sufficient such  cancellation would be easy to achieve in the case where some pair of regions has vanishing mutual information, namely when $I(A:C)=0$ or  $I(B:D)=0$.  In the former case, we can find a common maximizer flow for both regions by channeling the flow from\footnote{
In the interest of familiarity, here we conform to the convention of  \cite{Freedman:2016zud} in drawing the bit threads for a given region $X$ as directed {\it out of} $X$ into the bulk (rather than as spacetime-outward-directed flow used in \cite{Headrick:2017ucz}).
} 
$A$ into $BD$ but not $C$, and similarly channeling the flow from $C$ into $BD$ but not $A$.  In particular, when subadditivity for $A$ and $C$ is saturated, the flow bottleneck for $AC$ coincides with the union of the individual bottlenecks for $A$ and for $C$, which means that any maximizer flow $\v{AC}$ in fact saturates the norm bound on $\m{A}$ and on $\m{C}$, so that it is by definition simultaneously a maximizer flow for these individual regions, i.e.,
$\v{AC}=\v{A} = \v{C}$.
In such a case, using additionally the nesting of $B$ and $ABC$ to set $\v{ABC}=\v{B}$, 
the expressions in \req{e:v123def} would simplify to 
\begin{equation}
\pv_\lc =   \v{B} \ , \ \ \
\pv_\la =  \v{B}  \ , \ \ \
\pv_\lb = \v{A} 
\qquad ({\rm if} \ I(A:C)=0 )
\label{e:flowAC}
\end{equation}	
which means that  all the $\pv_i$'s are manifestly flows.

In the alternate case of $I(B:D)=0$, we can similarly set $\v{B}=\v{D}=-\v{ABC}$.  Having used up $\v{ABC}$, one might worry that we can no longer use nesting to get a further simplification; however, as one can see from \fig{f:simple}, this case was in fact just a rotated/relabeled version of the previous case, so the same arguments should still go through.
 And indeed, we still do have a form of nesting: $A\subset ABD$, which along with purity $\overline{ABD} =C$ can be used to set $\v{A}=\v{ABD}=-\v{C}$.   In this case,
\req{e:v123def} would simplify to 
\begin{equation}
\pv_\lc =   \v{A} \ , \ \ \
\pv_\la =  -\v{A}  \ , \ \ \
\pv_\lb = -\v{B} 
\qquad ({\rm if} \ I(B:D)=0 )
\label{e:flowBD}
\end{equation}	
so once again, all the $\pv_i$'s are manifestly flows.

One might at this point be tempted to declare victory:  after all, \req{e:SAC} implies that at least  one of $I(A:C)$ and $I(B:D)$ vanishes, since if \req{e:BDneck} holds, then $S(AC)=S(A)+S(C)$,
whereas for the opposite sign of the inequality, $S(BD)=S(AC)=S(B)+S(D)$.

However, \req{e:SAC} was predicated on the simple setup sketched in \fig{f:simple}.  We can easily have both $I(A:C)$ and $I(B:D)$ non-zero (and even make both diverge) by taking all regions to be pairwise near (or even adjoining).\footnote{
In the AdS$_3$ case, this can be achieved by having some of the regions be composed of multiple intervals (as exemplified in \fig{f:multAs}), while in higher dimensions, this is possible even with a single simply-connected region per subsystem.}
Since we can generically have both $I(A:C)>0$ and $I(B:D) > 0$,  the previous argument using \req{e:flowAC} or \req{e:flowBD} would not apply.
On the other hand, we can still use nesting along with purification   in full generality: e.g.,  we could use nesting of $B \subset ABC$ to set $\v{B}=\v{ABC}$, and use nesting of $A\subset ABD = {\overline C}$ to set $\v{A}=-\v{C}$.  Unfortunately, substituting these into \req{e:v123def}  gives
\begin{equation}
\pv_\lc =  \v{B} + \v{A} \ , \ \ \
\pv_\la =  \v{B}  - \v{A}  \ , \ \ \
\pv_\lb = 0\ , 
\label{e:nonflow}
\end{equation}	
so although we can achieve full cancelation in one of the $\pv_i$'s, the other two fail to be flows since they exceed the norm bound.  Hence we need to find a more general method of constructing the requisite maximizer flows.

\subsection{Motivation for our strategy}
\label{ss:plan}

To motivate such a construction, let us then first consider the structure of a generic flow.
Since each bit thread has two endpoints, located somewhere on the boundary (i.e.\ within either $A$ or $B$ or $C$ or $D$), we can think of a most general flow as composed out of (a subset of) {\it \strands} joining all possible pairs of regions.  In particular, in general we could have \strnds\ connecting 
$A \leftrightarrow B$, 
$A \leftrightarrow C$, 
$A \leftrightarrow D$,
$B \leftrightarrow C$,
$B \leftrightarrow D$,
and  $C\leftrightarrow D$.
Being a piece of a flow, each \strand\ individually satisfies the flow conditions \req{e:flowdef}, and if we're decomposing a single flow, these pieces manifestly cannot overlap.  On the other hand, viewed as more abstract building blocks for a collection of flows, we could  in general consider overlapping \strands, as long as they uphold the total norm bound.\footnote{
In terminology of \cite{MattMMI}, such objects are dubbed `multiflows'.  In the present work (unlike \cite{MattMMI}), we will in fact construct \strands\ which do not overlap.
}
To use the \strands\ as building blocks of a flow for a given region $X$, 
we simply consider all \strands\ emanating from $X$.\footnote{
We could optionally also include other \strands\ connecting pairs of regions distinct from $X$ (and not interfering with $X$'s \strnds), but since these are irrelevant for computing $S(X)$, we can choose to ignore them.
}

So far we have merely refined the nomenclature for a given flow, but we will now explain the utility of considering such a decomposition. Recall that flows for  non-nested but correlated regions $\v{A}$ and $\v{B}$ are incompatible.  This however does not necessarily mean that their individual \strands\ are likewise.  In particular, one can envision constructing all four maximizer flows, $\v{A}$, $\v{B}$, $\v{C}$, and $\v{D}=-\v{ABC}$ out of the {\it same} set of pre-specified \strands, by picking their orientations appropriately.  For example the \strnd\ $A \leftrightarrow B$ would contribute to $\v{A}$ in the orientation $\v{A\fto B}$ and  to $\v{B}$ in the opposite orientation $\v{B\fto A} = -\v{A\fto B} $.  (Note that by purity and nesting, we can always use the same set of threads to obtain flows for two disjiont regions $X$ and $Y\subset \overline{X}$ by taking $\v{X} = -\v{\overline{X}} = -\v{Y}$.)
As is intuitively clear and easy to check explicitly, keeping all the \strands\ spatially disjoint from each other would then guarantee that each of $\v{A}$, $\v{B}$, $\v{C}$, and $\v{D}$ constructed from them is likewise a flow.  Moreover, each is a maximizer flow if the \strands\ collectively render each of  $\m{A}$, $\m{B}$, $\m{C}$, and $\m{D}$ to be everywhere crossed perpendicularly by some \strnd\ (with the correct orientation)  with unit norm.

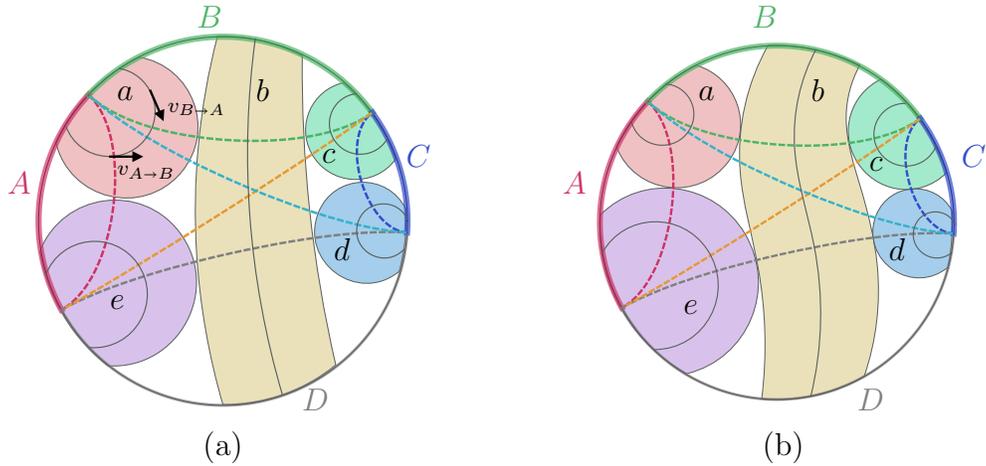
\begin{figure}
\begin{center}
\includegraphics[width=2.1in]{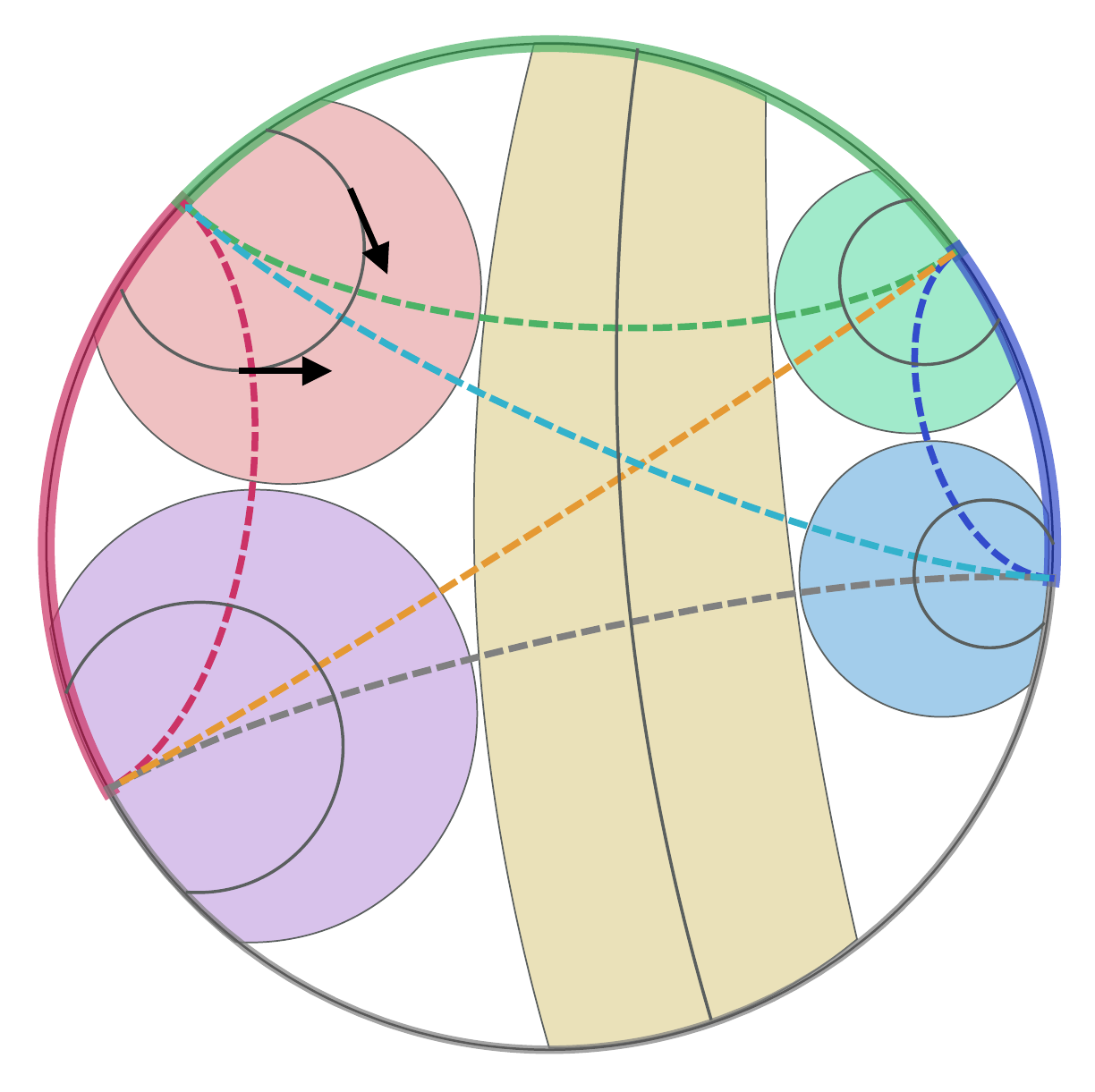}
\hspace{1.75cm}
\includegraphics[width=2.1in]{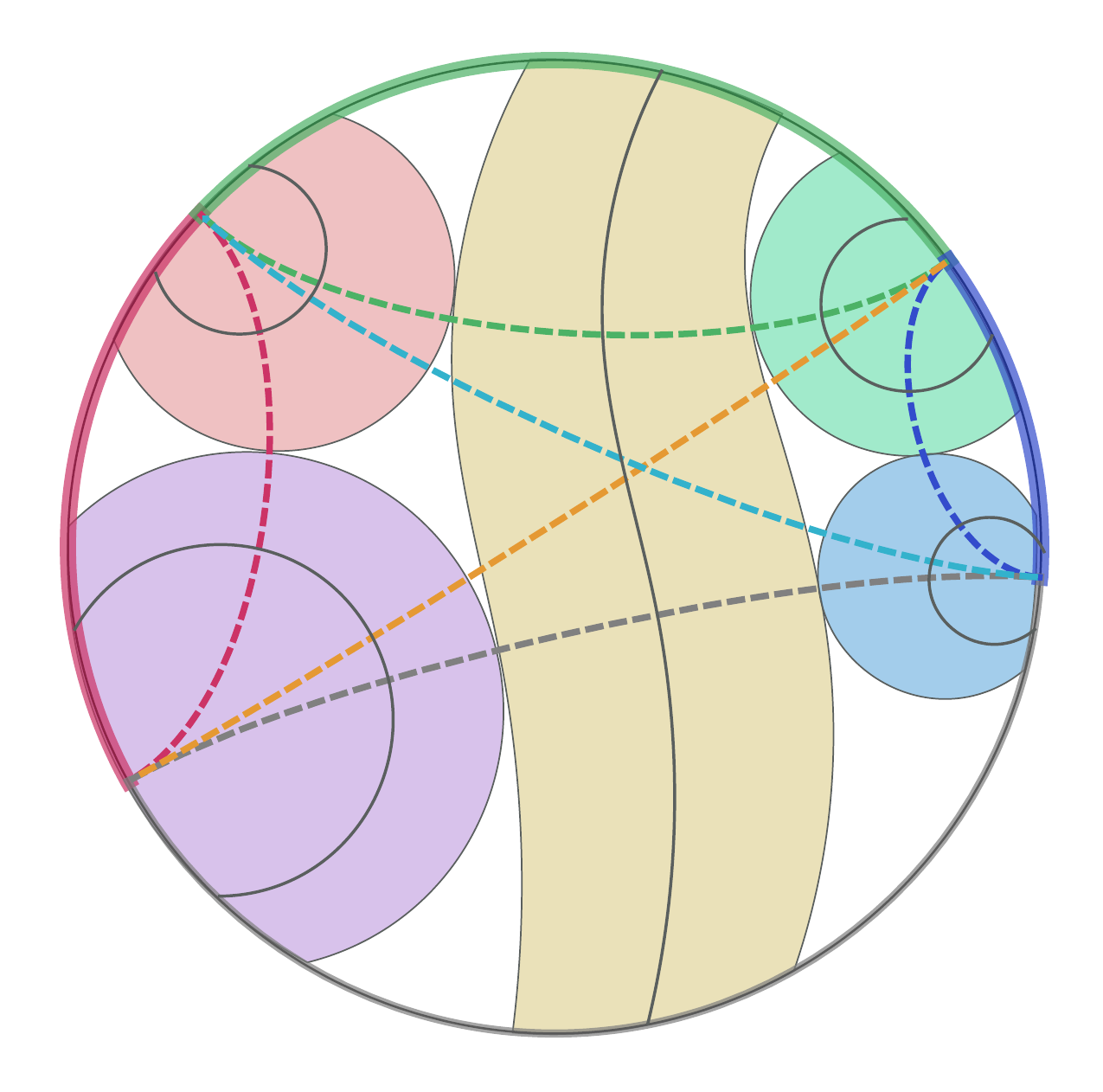}
\put(-162,80) {\makebox(20,20) {{\color{Acol}{$A$}}}}
\put(-90,143) {\makebox(20,20) {{\color{Bcol}{$B$}}}}
\put(-11,90) {\makebox(20,20) {{\color{Ccol}{$C$}}}}
\put(-50,-2) {\makebox(20,20) {{\color{Dcol}{$D$}}}}
\put(-372,80) {\makebox(20,20) {{\color{Acol}{$A$}}}}
\put(-300,143) {\makebox(20,20) {{\color{Bcol}{$B$}}}}
\put(-221,90) {\makebox(20,20) {{\color{Ccol}{$C$}}}}
\put(-260,-2) {\makebox(20,20) {{\color{Dcol}{$D$}}}}
\put(-324,84) {\makebox(20,20) {\scriptsize $\v{A\fto B}$}}
\put(-305,108) {\makebox(20,20){\scriptsize {$\v{B\fto A}$}}}
\put(-332,114) {\makebox(20,20) {$a$}}
\put(-280,115) {\makebox(20,20) {$b$}}
\put(-255,90) {\makebox(20,20) {$c$}}
\put(-250,55) {\makebox(20,20) {$d$}}
\put(-335,35) {\makebox(20,20) {$e$}}
\put(-112,114) {\makebox(20,20) {$a$}}
\put(-70,115) {\makebox(20,20) {$b$}}
\put(-48,87) {\makebox(20,20) {$c$}}
\put(-40,55) {\makebox(20,20) {$d$}}
\put(-118,33) {\makebox(20,20) {$e$}}
\vspace{0.4cm}
\put(-285,-10) {\makebox(0,0) {(a)}}
\put(-73,-10) {\makebox(0,0) {(b)}}
\caption{Separate components ({\it \strands}) of the maximizer flows  $\v{A}$, $\v{B}$, $\v{C}$, and $\v{D}$, passing through separate parts of the corresponding minimal surfaces as indicated, though only a single thread for each \strnd\ is shown explicitly.    Each \strand\ is confined to its own spatial region, which we compactly label $a$ (for  $A \leftrightarrow B$  \strnd), $b$ (for $B \leftrightarrow D$ \strnd), $c$ (for $B \leftrightarrow C$ \strnd),  $d$ (for $C\leftrightarrow D$ \strnd), and $e$ (for $A \leftrightarrow D$ \strnd).   The two panels illustrate a residual 1-parameter freedom in the construction, further discussed in \sect{sss:flowpartitions}.   {\bf (a)} In the $a$ \strnd, the arrows indicate that the same thread can be used in both directions, labeled $\v{A\fto B}$ and $\v{B\fto A}$. 
{\bf (b)} The relative sizes of the \strands\ are not fixed; one can vary them at the expense of e.g.\ bending the $b$ flow.
 }
\label{f:flowregions}
\end{center}
\end{figure}

The most economical way to construct disjoint \strands\ is to require that they in fact saturate the norm bound everywhere within their domain of support, since accommodating a smaller norm for any \strnd\ would require greater bulk volume. 
In other words, we wish to comb the flow configuration into maximally collimated \strands, in such a way that these \strnds\ do not intersect each other.
Consequently, for our configurations (cf.\ \fig{f:simple}), the presence of an $A \leftrightarrow C$ \strnd\ is incompatible with simultaneous presence of a $B \leftrightarrow D$ \strnd.  This is however not a problem, since as explained above, $I(A:C)=0$ obviates the need for the  $A \leftrightarrow C$ \strnd.\footnote{
This is manifest by choosing $\v{A}=\v{C}$, but our construction will actually be more thrifty, and have no overlap at all between the two flows (i.e., implement $|\v{A}| \, |\v{c}| = 0$ everywhere).  Similarly, the alternate case $I(B:D)=0$ would obviate the need for the  $B \leftrightarrow D$ \strand.}
 This reasoning motivates us to try constructing a \strand\ configuration of the form sketched in \fig{f:flowregions}, which exemplifies how each maximizer flow can be partitioned into several \strands, each remaining within its associated bulk region; for compactness of notation we label these $a,b,c,d,e$ as indicated in \fig{f:flowregions}. 

Once we have constructed such a set of \coop \strands\ from which we can assemble all four maximizer flows, it follows almost immediately that their linear combinations  $\pv_\la$, $\pv_\lb$, and  $\pv_\lc$ defined via \req{e:v123def}
will likewise be flows.  The crux of our MMI proof then boils down to showing that a configuration envisioned in \fig{f:flowregions} is indeed always tenable, which we will now demonstrate by an explicit construction.

\subsection{Flow construction}
\label{ss:constr}

In this section we present an explicit construction for the maximizer flows 
$\v{A}$, $\v{B}$, $\v{C}$, and $\v{ABC}$ 
which render the $\pv_i$'s of  \req{e:v123def} likewise flows.  

Starting with partitioning of AdS$_3$ boundary time slice of the type sketched in \fig{f:simple},
 we first show that we can partition the maximizer flows as indicated in Table \ref{t:MFsplit} (cf.\ \fig{f:flowregions}).
\begin{table}[htp]
\caption{Partitioning of the 4 maximizer flows (indicated in the left column) into \strands\ confined to the 5 bulk regions (indicated in the top row).  Each row corresponds to an equation delineating this partition, e.g.\ the first row gives $\v{A} = \v{A\fto B} + \v{A\fto D}$, where the two RHS terms are supported entirely within the regions $a$ and $e$, respectively, and similarly for the other rows.}
\begin{center}
\begin{tabular}{|c|c|c|c|c|c|}
& $a$ & $b$ & $c$ & $d$ & $e$ \\
\hline \hline
$\v{A}$ &  $\v{A\fto B}$ & 0 & 0 & 0 & $\v{A\fto D}$ \\
\hline
$\v{B}$ & $\v{B\fto A}$ & $\v{B\fto D}$  & $\v{B\fto C}$ & 0 & 0 \\
\hline
$\v{C}$ & 0 & 0 &  $\v{C\fto B}$ & $\v{C\fto D}$  & 0 \\
\hline
$\v{ABC}$ & 0 & $\v{B\fto D}$ & 0 & $\v{C\fto D}$ & $\v{A\fto D}$ \\
\hline
\end{tabular}
\end{center}
\label{t:MFsplit}
\end{table}
In particular, in \sect{sss:flowpartitions} we argue that we can comb the maximizer flow from $A$ into a \strand\footnote{
Recall that for $X,Y$ chosen from the fundamental regions $A$, $B$, $C$, and $D$, the notation $\v{X \fto Y}$ simply indicates the part of the $\v{X}$ which flows into $Y$.  The only requirement that we impose on such a partitioning is that $\v{X \fto Y} = -\v{Y \fto X}$ for all $X,Y$.  Since this corresponds to merely reversing the orientation of the bit threads, both flows have the same norm and divergence.  
}  $\v{A\fto B}$ leading to $B$ and a \strand\ $\v{A\fto D}$ leading to $D$ (but not an  $\v{A\fto C}$ \strand\ thanks to \req{e:BDneck}), and similarly for $B$, $C$, and $D$, such that the \strands\ pass only through restricted {\it disjoint} bulk regions (denoted $a,b,c,d,e$).

Once we show (in the remainder of this section) that such a partition is possible, the rest is straightforward:  Applying the definition \req{e:v123def}, we can directly evaluate the $\pv_i$'s, and examine them within each region.  The result is packaged in Table \ref{t:PFcheck}, which gives the explicit contributions within each region.
\begin{table}[htp]
\caption{Explicit check that the $\pv_i$'s defined by \req{e:v123def} are indeed flows, given the partition in Table \ref{t:MFsplit}.  Each row gives an equation with LHS being the first column and RHS the sum of the terms in the remaining columns, as in Table \ref{t:MFsplit}.  The columns correspond to the distinct bulk regions (cf.\ \fig{f:flowregions}) and we see that in each region, $\pv_i$ is manifestly a flow. }
\begin{center}
\begin{tabular}{|c|c|c|c|c|c|}
& $a$ & $b$ & $c$ & $d$ & $e$ \\
\hline \hline
\  $\pv_\lc$ \quad & 0 & $\v{B\fto D}$ & $\v{B\fto C}$ & 0 & $\v{A\fto D}$  \\
\hline
\  $\pv_\la$ \quad & $\v{B\fto A}$ &  $\v{B\fto D}$ & 0 &$\v{C\fto D}$  & 0 \\
\hline
\  $\pv_\lb$ \quad & $\v{A\fto B}$ & 0 & $\v{C\fto B}$  & $\v{C\fto D}$ & $\v{A\fto D}$  \\
\hline
\end{tabular}
\end{center}
\label{t:PFcheck}
\end{table}
The crucial observation here is that each entry in the table is of the form $\v{X \fto Y}$ (i.e.\ a \strnd\ of a single flow $\v{X}$ and therefore manifestly a flow).  Hence even though each of the $\pv_i$'s is the sum of three bulk flows, the result is nevertheless of unit-bounded norm everywhere.  This establishes that $\pv_i = v_i$ (i.e.\ each $\pv_i$ is a flow) as advertised.\footnote{
\label{fn:IACgen}
Recall that the above construction assumed that $I(A:C)=0$ which implies \req{e:BDneck}.  However, as explained in footnote \ref{fn:IAC}, this was merely a notational convention.  If $I(A:C)>0$ so that the $BD$ flows have more stringent bottleneck than the $AC$ system, we can simply reconfigure the \strands\ so as to have a component  $\v{A\fto C}$ instead of $\v{B\fto D}$, along with the corresponding rearrangement of the regions $a,b,c,d,e$ (such that $a$ adjoins $c$, $d$ adjoins $e$, and $b$ passes between $a$ and $e$ and between $c$ and $d$).  Consequently in Table \ref{t:PFcheck}, the $b$ column would contain $\v{A\fto C}$ and $\v{C\fto A}$ in the first and 2nd rows respectively, while the rest of the table remains identical.
}

\subsubsection{Maximizer flow partitions}
\label{sss:flowpartitions}

We now explain why the partition indicated in Table \ref{t:MFsplit} and exemplified in  \fig{f:flowregions} is indeed always possible.  
We have already seen that the decomposition summarized in the rows of Table \ref{t:MFsplit} without requiring the bulk regions $a$, $b$, $c$, $d$, and $e$ to be disjoint,  is necessitated by the divergencelessness condition \req{e:flowdef} (since each thread from a given region has to end in one of the other regions), along with the observation that we need not activate any \strands\ for uncorrelated regions.
What remains to be shown is the compatibility between these respective partitions, namely that the \strands\ are mutually non-overlapping throughout the bulk.
Since a general maximizer flow from each region necessarily saturates the norm bound on the associated minimal surface presenting a bottleneck for the flow, the first step is to verify the consistency of partitioning on  the corresponding minimal surfaces.  
 
 In the present case, the surface $\m{A}$ splits into two parts, $\mm{A}{a}$ and $\mm{A}{e}$, associated with the \strands\ ($a$ and $e$) ending on $B$ and $D$, respectively.  Similarly, $\m{B}$ splits into three parts, $\mm{B}{a}$, $\mm{B}{b}$, and $\mm{B}{c}$, channeling the \strnds\ ($a,b,c$) from $B$ into $A$, $D$, and $C$, respectively, and so on.   
Moreover, since the threads of $\v{X \fto Y}$ through $\mm{X}{z}$ coincide with the threads of $\v{Y \fto X}$ through $\mm{Y}{z}$, but with opposite orientation, the areas of the two minimal surface parts must likewise match,  $\area{\mm{X}{z}} = \area{\mm{Y}{z}}$, for all $X,Y \in {A,B,C,D}$ and corresponding $z \in {a,b,c,d,e}$. 

Written out explicitly, given
$\m{A}$, $\m{B}$, $\m{C}$, and $\m{D}$, the minimal surface subdivisions need to satisfy the following constraints: 
\begin{equation}
\begin{split}
\area{\m{A}}  & = \area{\mm{A}{a}} +\area{\mm{A}{e}} \\
\area{\m{B}}  & = \area{\mm{B}{a}} +\area{\mm{B}{b}} +\area{\mm{B}{c}}  \\
\area{\m{C}}  & = \area{\mm{C}{c}} +\area{\mm{C}{d}}  \\
\area{\m{D}}  & = \area{\mm{D}{e}} +\area{\mm{D}{b}} +\area{\mm{D}{d}} \\
& \area{\mm{A}{a}} = \area{\mm{B}{a}} \\
& \area{\mm{B}{b}} = \area{\mm{D}{b}} \\
& \area{\mm{B}{c}} = \area{\mm{C}{c}} \\
& \area{\mm{C}{d}} = \area{\mm{D}{d}} \\
& \area{\mm{A}{e}} = \area{\mm{D}{e}} 
\end{split}
\label{e:areaconds}
\end{equation}
This constitutes 9 equations for 10 unknowns, and therefore we expect to have a 1-parameter family of solutions, as illustrated in \fig{f:flowregions}a and \fig{f:flowregions}b.  One procedure of constructing a partitioning is as follows.  Choose the partition of $\m{A}$ (in the present case that is just a single point $p_A$).  This determines $\mm{A}{a}$ and $\mm{A}{e}$, and hence also $\mm{B}{a}$ and $\mm{D}{e}$.  Now find the partition of $\m{C}$ (determined by a point $p_C$, which immediately specifies $\mm{C}{c}$, $\mm{B}{c}$, $\mm{C}{d}$, and $\mm{D}{d}$) so as to satisfy the remaining equation of \req{e:areaconds}, namely $\area{\mm{B}{b}} = \area{\mm{D}{b}}$.  Since if we slide $p_C$ sufficiently ``up'' along $\m{C}$ towards $B$, we close off $\mm{D}{b}$, whereas if we slide $p_C$ ``down'' towards $D$, we close off $\mm{B}{b}$, a solution clearly exists where the two terms become equal.  This completes the argument that minimal surface partition envisioned in \fig{f:flowregions}, i.e.\ satisfying \req{e:areaconds}, always exists.

Now that we have confirmed that a viable partitioning of the 4 minimal surfaces exits, we discuss the procedure for constructing the full maximizer flow for each region.  
There is of course large freedom in how we construct the requisite maximizer flows; here we present two particularly simple constructions which are easy to generalize.  Although the flows are required to saturate the norm bound only on the associated minimal surface, we will find it convenient to construct them so as to remain maximally packed (and therefore equidistant) everywhere, since this utilizes the bulk geometry most economically.  
Geometrically, such a construction is actually quite simple, and is based on the observation explained in \cite{Headrick:2017ucz} that a foliation of a bulk region by minimal surfaces induces a maximally-collimated flow.  Intuitively this is because the minimal surfaces have by definition vanishing expansion for the normal congruence (i.e., vanishing extrinsic curvature), so the bit threads which pass perpendicularly through any leaf of such foliation locally cannot change their cross-sectional density; in other words, they must remain maximally packed across any minimal surface they cross perpendicularly.\footnote{
One might also naively worry that the flows encounter other bottlenecks which we have not taken into account.  For example the $b$ \strand\ crosses both $\m{AB}$ and $\m{BC}$ (cf.\ \fig{f:flowregions}), which a-priori need not have larger area than, say, $\m{B}$.  However, this is not a problem (as we explicitly demonstrate below), since the non-perpendicularly-crossed surfaces have by construction smaller thread density and therefore larger area.
}

\subsubsection{Foliation building blocks}
\label{sss:foliations}

One strategy to obtain maximally packed \strand, then, is to start by considering a family of minimal surfaces which foliate the bulk space, since the normal congruence to these surfaces describes a flow.  
  In the present case of 2-dimensional Poincare disk geometry, the foliating surfaces are simply geodesics, which indeed foliate the space as long as their endpoints vary continuously in any manner which generates a boundary foliation (which we can think of as a family of nested intervals whose endpoints collectively cover the boundary).\footnote{
Recall that bulk minimal surfaces (here geodesics) anchored on nested regions (here intervals) are guaranteed not to intersect (since doing so would contradict global minimality), and the AdS geometry guarantees that there is a unique geodesic for each boundary interval and that a geodesic through any point and in any direction has both its endpoints on the boundary.  The latter two properties do not hold universally for arbitrary static asymptotically AdS geometries, though they do hold robustly when the deformations from pure AdS are not too large.  We revisit the more general case in \sect{ss:geom}.
}  Labeling each leaf of the foliation by a parameter $\lambda\in [0,1]$, we can specify the foliation by  two functions, $\ep^L(\lambda)$ and $\ep^R(\lambda)$, corresponding to the angular position of the two endpoints of the boundary interval.  A foliation requires each function to be monotonic, but in opposite directions (so as to get nested intervals), such that  $\ep^L(0)=\ep^R(0)$ and $\ep^L(1)=\ep^R(1)+ 2\pi$ .  

\begin{figure}[htbp]
\begin{center}
\includegraphics[width=1.6in]{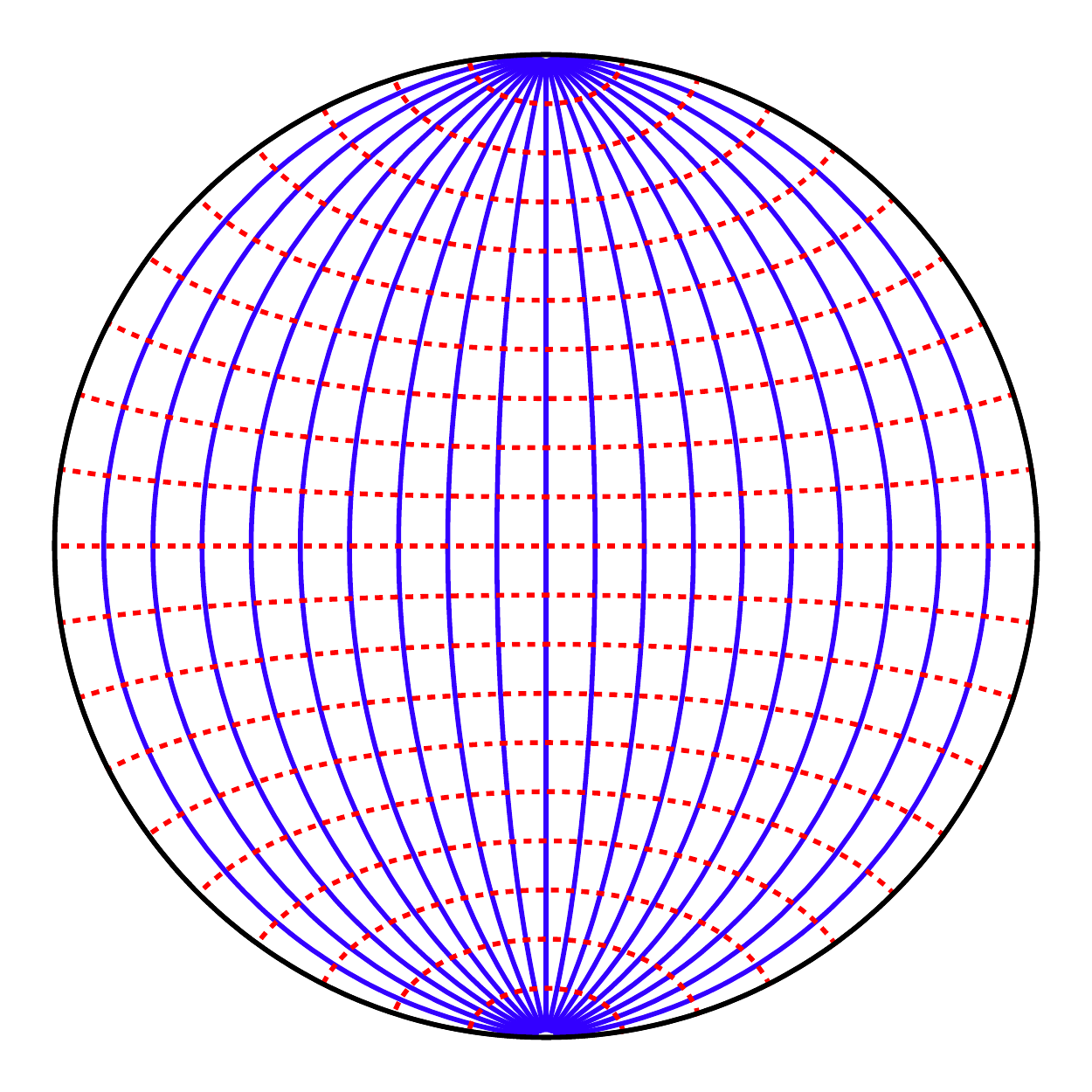}
\hspace{.5cm}
\includegraphics[width=1.6in]{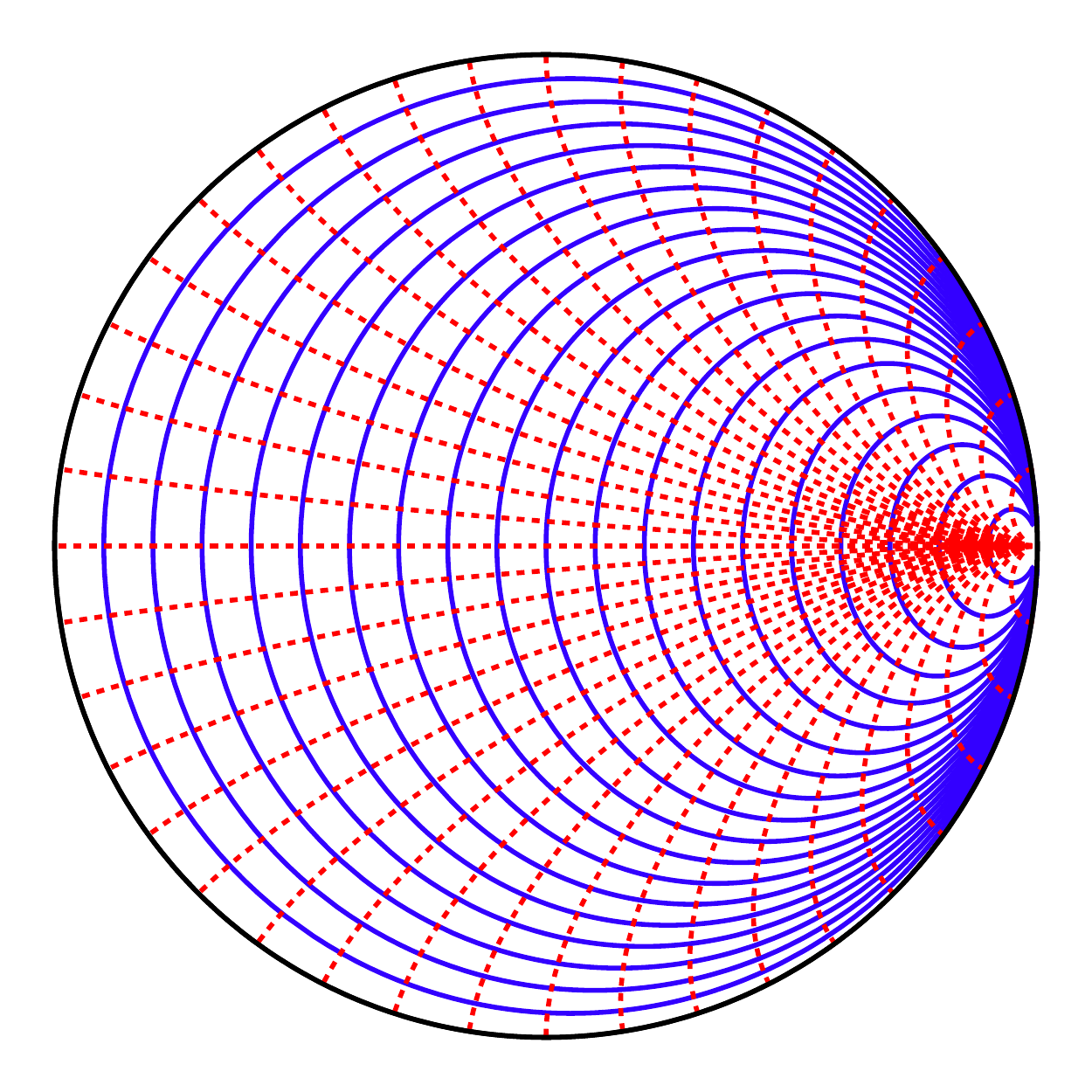}
\hspace{.5cm}
\includegraphics[width=1.6in]{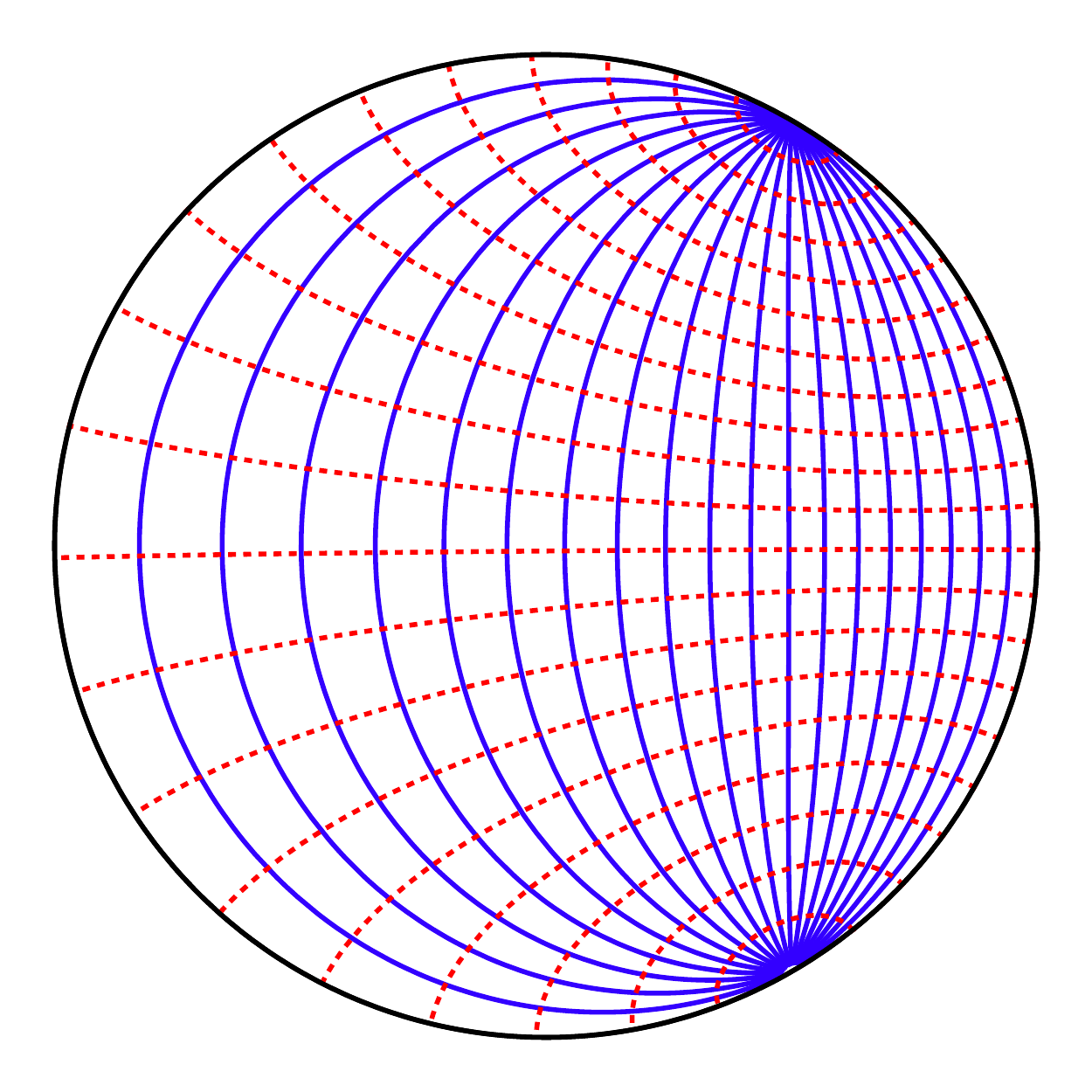}
\vspace{0.4cm}
\put(-337,-10) {\makebox(0,0) {(a)}}
\put(-200,-10) {\makebox(0,0) {(b)}}
\put(-62,-10) {\makebox(0,0) {(c)}}
\caption{Examples of flows (solid blue curves) induced by minimal surface foliations (dotted red curves).  (While we use the coordinates of \fig{f:simple}, we don't maintain constant proper spacing  between the treads -- doing so would clump them too much near the boundary.)  Such foliations are specified by the endpoints $\ep$ of the bounding interval, as functions of $\lambda \in [0,1]$ (see text for details).  In particular, we illustrate foliations generating
{\bf (a)} antipodal threads, 
{\bf (b)} threads with both endpoints pinned at a single point of the Poincare disk boundary, and
{\bf (c)} threads with endpoints pinned at generically-separated points.
}
\label{f:foliations}
\end{center}
\end{figure}

We sketch some simple examples in \fig{f:foliations}.
In \fig{f:foliations}a, the endpoints vary symmetrically, e.g.\ $\ep^{L,R}(\lambda) = \frac{\pi }{2} \pm \pi \, \lambda$.  In this case the threads happen to coincide with constant-time projections of null geodesics through AdS.  In \fig{f:foliations}b, we fix one endpoint and let the other span the rest of the boundary, $\ep^R(\lambda) =0, \ep^L(\lambda) = 2 \pi \, \lambda$.  In this case the threads happen to be horocycles on  the Poincare disk.  Although this looks like all the bit threads start and end at the same point, this is merely due to the conformal rescaling.  The flow lines actually follow constant-$z$ contours in Poincare coordinates $ds^2 = (-dt^2+dx^2+dz^2)/z^2$ (they are generated by geodesics at constant $x$), where it is manifest that they straddle the $x=-\infty$ and $x=+\infty$ parts of the boundary.\footnote{  
One can also resolve this confluence by working at finite cutoff, where we could either only require the endpoints to get within the cutoff scale from each other (e.g.\ $\ep^L(0)=\veps $ and $\ep^L(1)=2\pi - \veps $), or determine the endpoints by cutting off the same set of spacelike geodesics at finite radius (in which case both endpoints would vary).
}
One can likewise separate the flow ends, as shown in \fig{f:foliations}c.  For example for the endpoints given by $\ep^R(\lambda) = \ph_0- 2 \ph_0 \, \lambda$ and $\ep^L(\lambda) = \ph_0 +( 2 \pi -  2 \ph_0)  \lambda$, we see that this case interpolates between the first two.  
Of course, one can also have more irregular flows, and in fact all these flows can be patched together across any common minimal surface (such at the horizontal bisectors in \fig{f:foliations}).

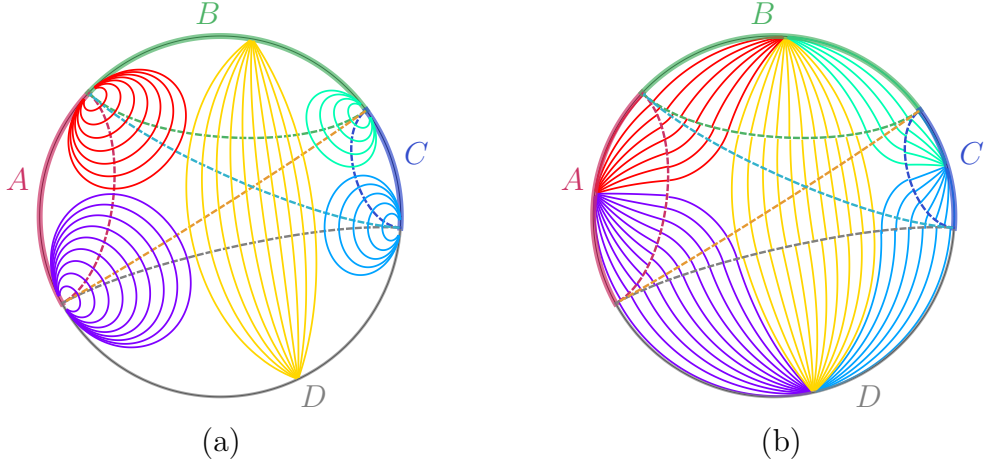
\begin{figure}[htbp]
\begin{center}
\includegraphics[width=2.1in]{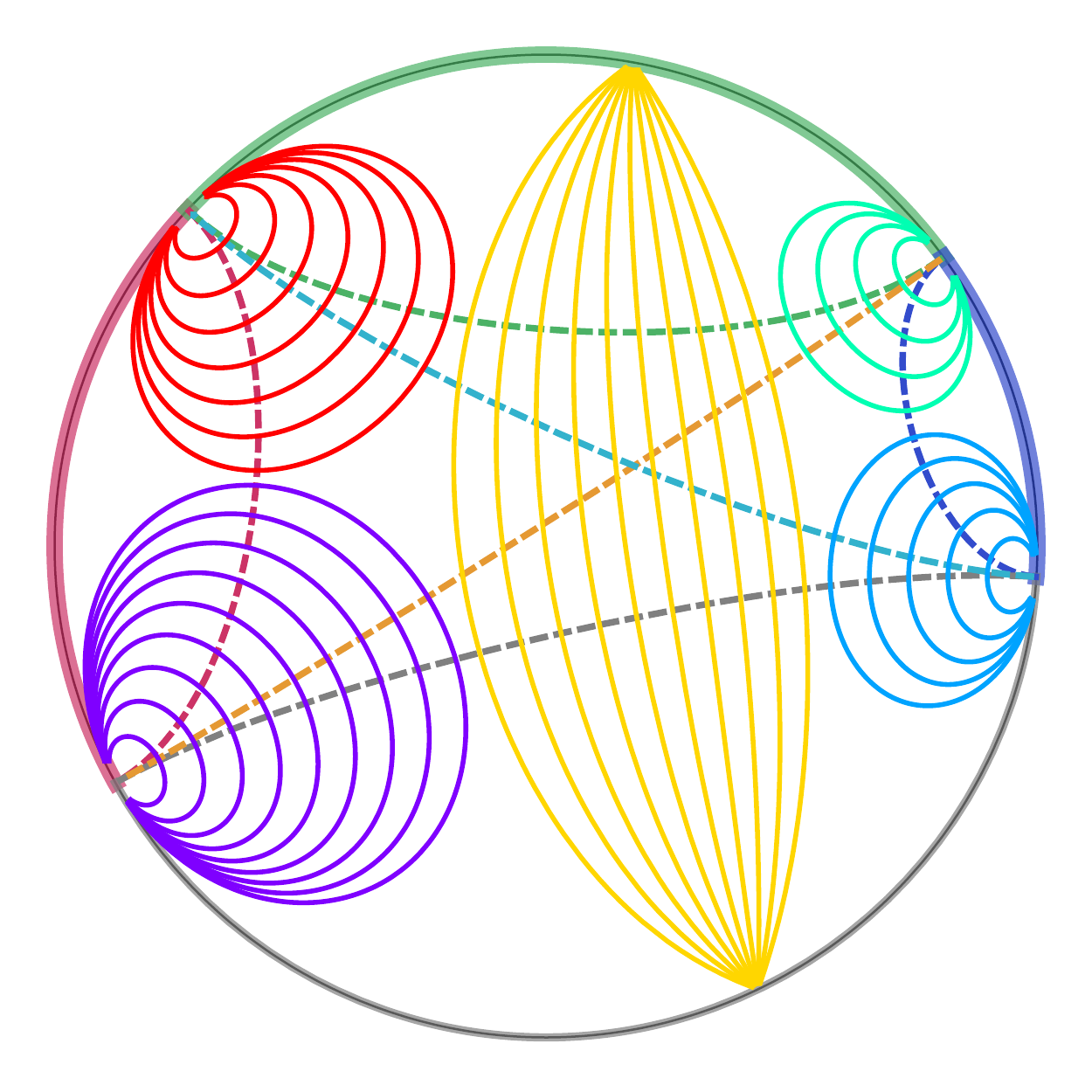}
\hspace{1.75cm}
\includegraphics[width=2.1in]{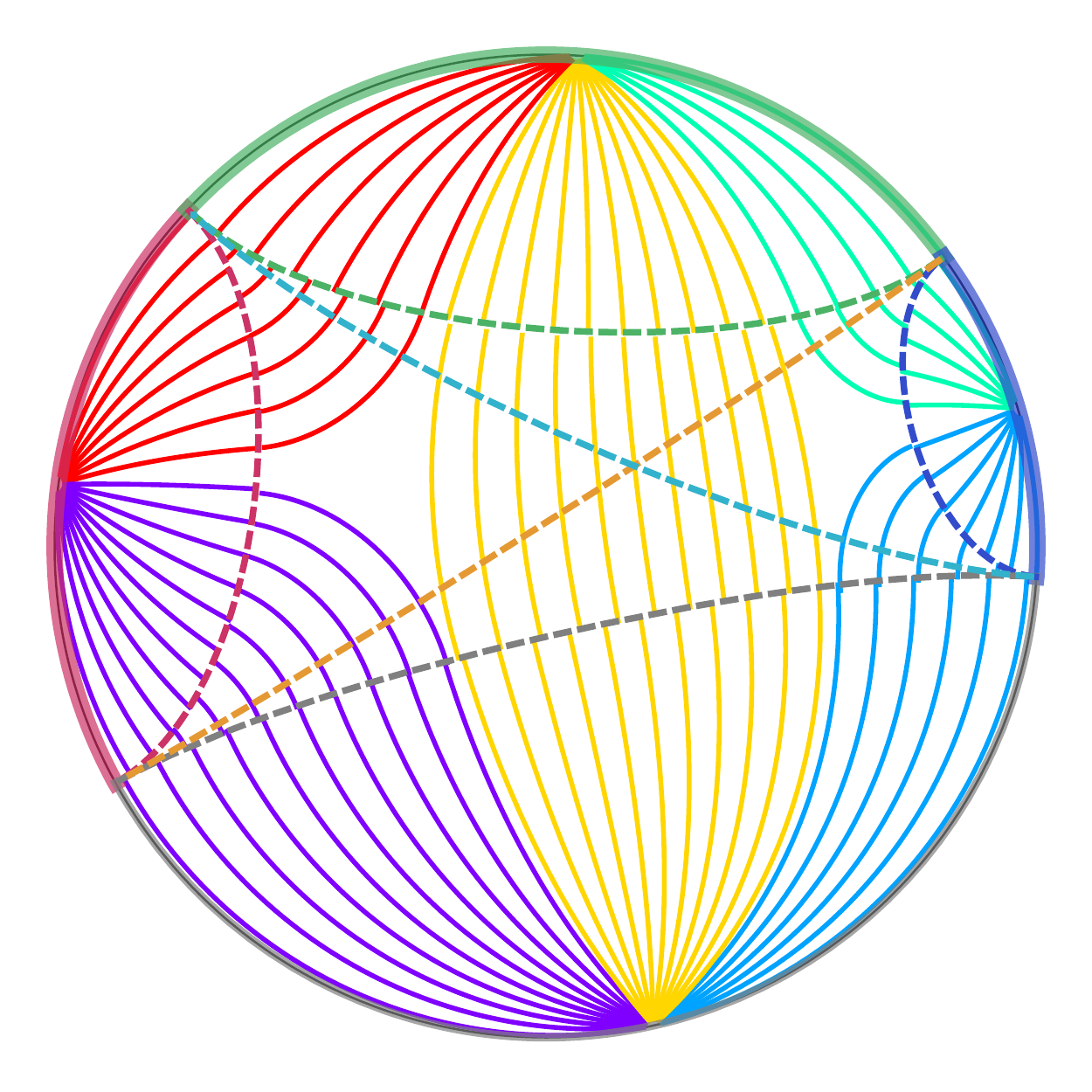}
\put(-162,80) {\makebox(20,20) {{\color{Acol}{$A$}}}}
\put(-90,143) {\makebox(20,20) {{\color{Bcol}{$B$}}}}
\put(-11,90) {\makebox(20,20) {{\color{Ccol}{$C$}}}}
\put(-50,-2) {\makebox(20,20) {{\color{Dcol}{$D$}}}}
\put(-372,80) {\makebox(20,20) {{\color{Acol}{$A$}}}}
\put(-300,143) {\makebox(20,20) {{\color{Bcol}{$B$}}}}
\put(-221,90) {\makebox(20,20) {{\color{Ccol}{$C$}}}}
\put(-260,-2) {\makebox(20,20) {{\color{Dcol}{$D$}}}}
\vspace{0.4cm}
\put(-285,-10) {\makebox(0,0) {(a)}}
\put(-73,-10) {\makebox(0,0) {(b)}}
\caption{Illustration of two distinct constructions for obtaining combed maximizer flows using foliations. {\bf (a)} smooth flows obtained by 5 superimposed foliations.  {\bf (b)} patched flows obtained by a single bulk `\locfol'.  The boundary regions and associated minimal surfaces are as in \fig{f:simple}, and the threads are indicated by solid curves, with distinct colors delineating the regions $a,b,c,d,e$ of \fig{f:flowregions} which contain the corresponding \strands.}
\label{f:flowconstructs}
\end{center}
\end{figure}

As we see, we can comb the flows in a large (continuously infinite) variety of ways, and from any boundary point to any other boundary point.  However,  such foliations cannot admit multiple \strands, namely multiple starting and ending points for a given flow.  To incorporate this more general case, we have to use multiple foliations simultaneously, or generalize the foliating surfaces to a more localized construct.  As an example of each, we consider two particularly natural constructions, shown in \fig{f:flowconstructs}.

\begin{enumerate}
\item
Use a complete smooth foliation (by entire minimal surfaces) for each \strand.  This leads to all threads being smooth, but a given foliation is `active' (in terms of guiding a thread) in only a subset of the bulk.  Hence there are multiple simultaneous bulk foliations.  See \sect{sss:FirstConstr} below for explicit construction and \fig{f:flowconstructs}a for the resulting plot.
\item
Use a single bulk foliation for (almost) the entire bulk, but by only piecewise-minimal surfaces.  The resulting flows are typically not smooth.
See \sect{sss:SecondConstr} for explicit construction and \fig{f:flowconstructs}b for the corresponding plot.
\end{enumerate}

\subsubsection{Foliation for each \strand, smooth flows}
\label{sss:FirstConstr}

\begin{figure}[htbp]
\begin{center}
\includegraphics[width=1.6in]{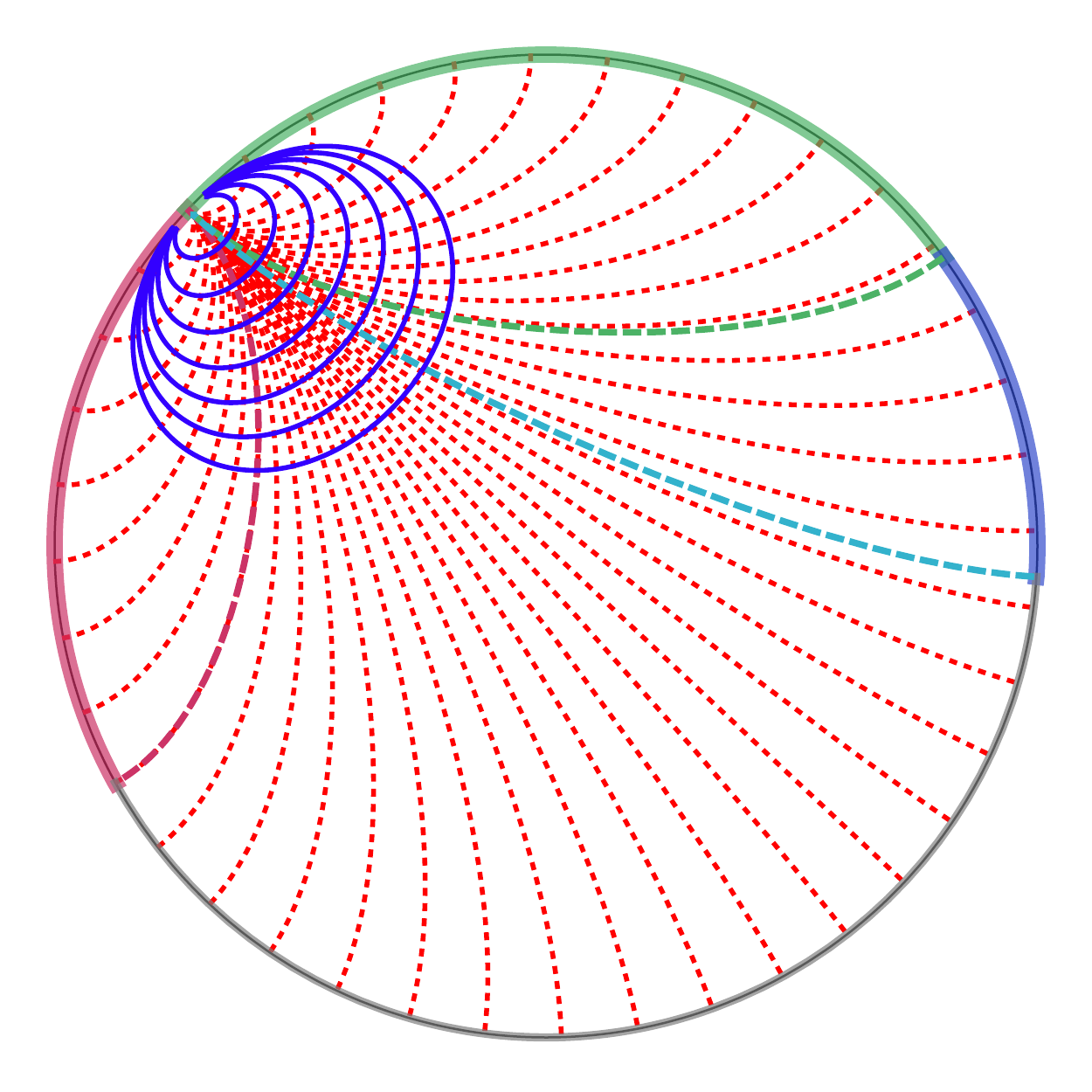}
\hspace{.5cm}
\includegraphics[width=1.6in]{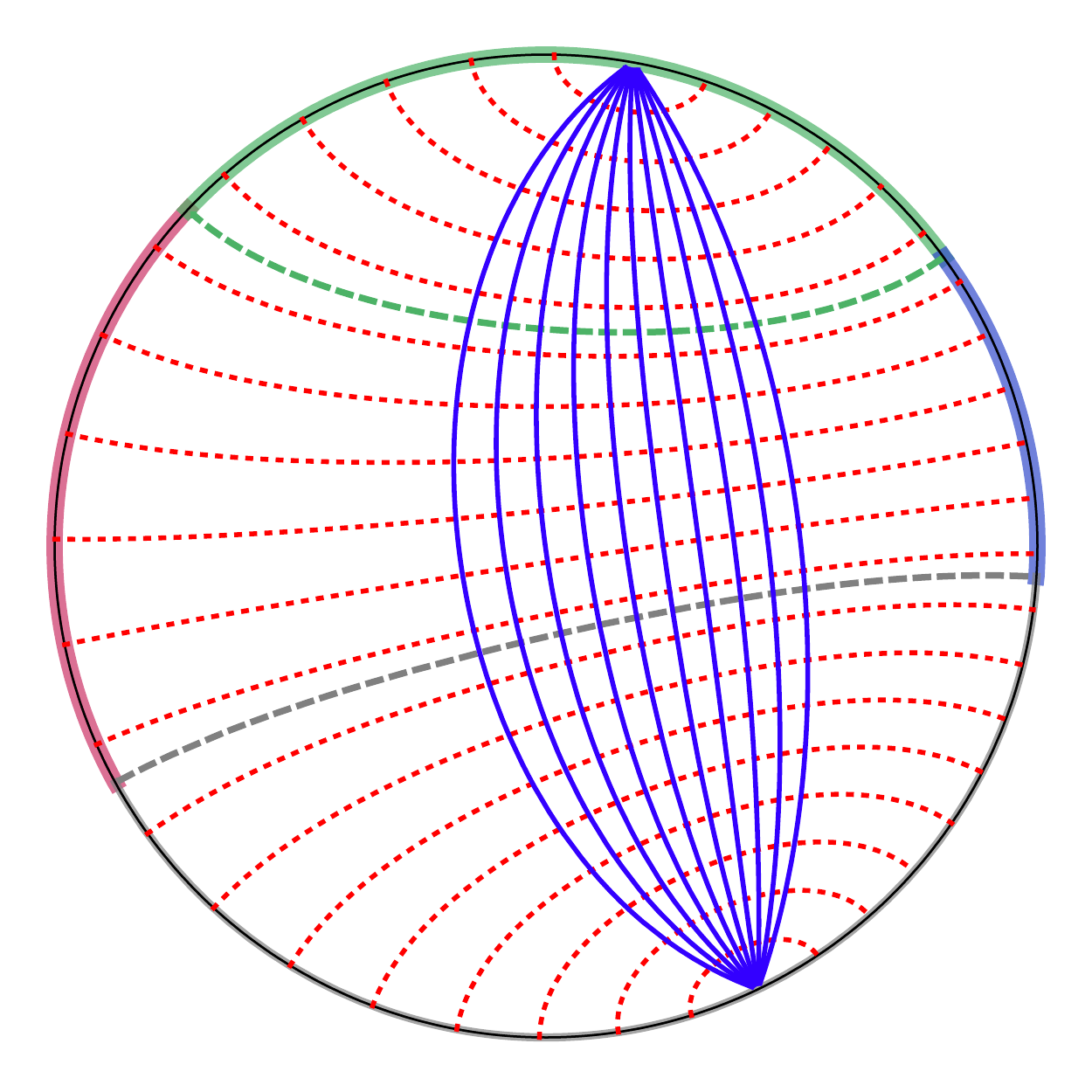}
\hspace{.5cm}
\includegraphics[width=1.6in]{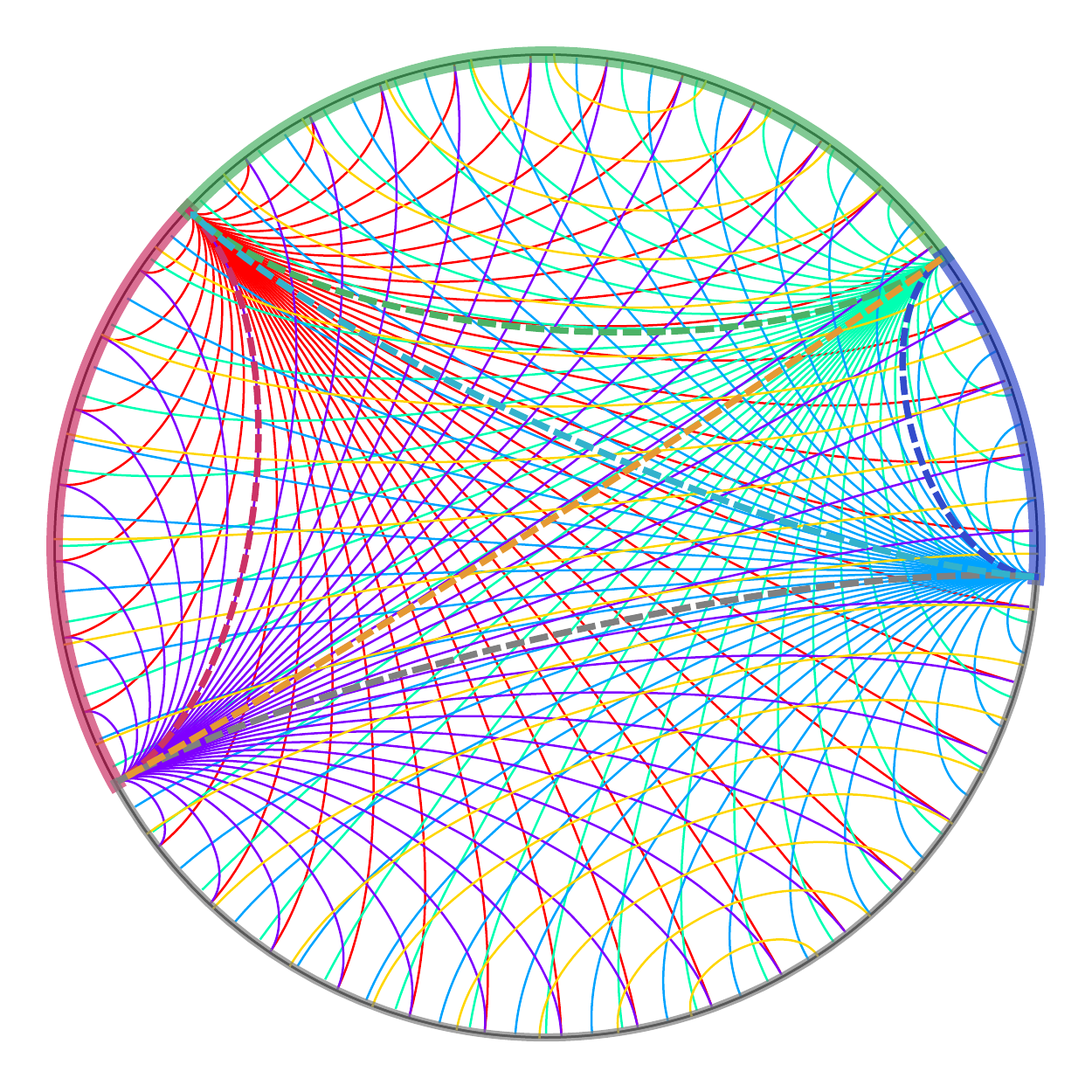}
\vspace{0.4cm}
\put(-337,-10) {\makebox(0,0) {(a)}}
\put(-200,-10) {\makebox(0,0) {(b)}}
\put(-62,-10) {\makebox(0,0) {(c)}}
\caption{Specific foliations which generate the smooth flow  configuration indicated in \fig{f:flowconstructs}a, in particular generating {\bf (a)} the flow $\v{A \fto B}$   
and {\bf (b)} the flow $\v{B \fto D}$.
When suitably restricted (as indicated by the solid blue curves) these give the \strands\ $a$ and $b$, respectively. {\bf (c)} All 5 foliations (color-coded as in \fig{f:flowregions}) superposed on each other.}
\label{f:examplefols}
\end{center}
\end{figure}

We first consider the case with full bulk foliation per \strnd.  Since we have 5 \strands\ confined to 5 separate bulk regions $a,b,c,d,e$, we will have 5 distinct foliations. Each foliation is specified by a family of minimal surfaces (geodesics) characterized by their endpoints $\ep^L(\lambda)$ and $\ep^R(\lambda)$ for $\lambda\in [0,1]$, as indicated in \fig{f:examplefols}.
For the $a,c,d$, and $e$ \strnds\ (defining the associated bulk regions), the foliations are all of the horocycle type sketched in \fig{f:foliations}b (rotated in global coordinates): the minimal surfaces `fan out' from one endpoint, say $\ep^L$, 
with the other endpoint $\ep^R$ spanning the rest of the boundary, $\ph^R(\lambda) = \ph^ L+2\pi \, \lambda$. These families then contain the minimal surfaces with the given fixed endpoint: the $a$ family, shown in \fig{f:examplefols}a, in particular contains $\m{A}$ and $\m{B}$ (as well as $\m{BC}$), the $c$ family contains $\m{B}$ and $\m{C}$, and so forth.
 The final family, $b$, indicated in \fig{f:examplefols}b, is of the type sketched in \fig{f:foliations}c, varying both endpoints so as to contain $\m{B}$ and $\m{D}$.\footnote{
 While we saw that the `bottleneck' equations \req{e:areaconds} in general have a 1-parameter family of solutions (illustrated in the two panels of \fig{f:flowregions}), in the present case that parameter is used up by choosing  $\v{B \fto D}$ to be of the particular form plotted in \fig{f:foliations}c (generated by threads emanating radially outward from a shifted origin in Poincare coordinates and rotated in global coordinates), so that the $B$ and $D$ positions fix this solution uniquely, to the type sketched in \fig{f:flowregions}a as opposed to \fig{f:flowregions}b.
 }
In \fig{f:examplefols}c we superpose all five foliations (without including the corresponding threads), to show how are the individual leaves of the foliations related to each other.

Since for a given \strand\ only a part of its corresponding foliation is used, our remaining task is to show that the utilized regions are mutually compatible.  In particular for the $\v{A \fto B}$ flow associated to the  bulk region $a$ (\fig{f:examplefols}a), we only use the foliating minimal surfaces for normal congruence passing through $\m{A \fto B}$ and  $\m{B \fto A}$ as indicated, and similarly for the other \strands.  In \sect{sss:flowpartitions} we have argued that the minimal surfaces $\m{A}$, $\m{B}$, $\m{C}$, and $\m{D}$ can be partitioned in a compatible fashion, i.e.\ satisfying the bottleneck equations \req{e:areaconds}.  Hence what remains is to show that the rest of the \strnds\ can likewise accommodate themselves so that distinct \strands\ don't overlap each other.   

Let us consider, say, the compatibility of $a$ and $b$ flows (the arguments for the compatibility of the other flows being identical).
We want to show that $\v{B \fto A}$ and $\v{B \fto D}$ do not intersect.   The most `worrisome' regions are where the \strands\ approach closest to each other, i.e.\ just around the minimal surface $\m{B}$. Both flows are maximally constrained at $\m{B}$, and they pass normally through it (consistently with $\m{B}$ belonging to both foliation families,  $a$ and $b$, cf.\ \fig{f:examplefols}a and \fig{f:examplefols}b).  Since at $\m{B}$ itself we ensured that the flows are compatible by the $\m{B \fto A},\m{B \fto D}$ partitioning, to see that subsequently the \strands\ bend away from each other, let us consider a leaf of the $a$ and $b$ foliations which have a common right endpoint $\ep_a^R(\lambda_a) = \ep_b^R(\lambda_b)=p_\beta + \epsilon$, just `above' $\m{B}$. The corresponding left endpoints are distinct: $\ep_a^L(\lambda_a) = p_\alpha$, while $\ep_b^R(\lambda_b)=p_\alpha - \epsilon'$, with the $b$-leaf's defining interval lying within the $a$-leaf's interval (as the reader may try  to verify by a careful look at \fig{f:examplefols}c).  This means that the $a$ leaf bends further to the left than the $b$ leaf (recall that the leafs cannot intersect since otherwise they would contradict the assumption of minimality in their construction).  Therefore, their normals (which guide their respective flows) likewise bend away from each other: $n_a$ (pointing towards $B$) more to the left and $n_b$ more to the right.  We can make a similar argument for the behavior of the flow `below' $\m{B}$ where the flows likewise bend away from each other in the downward direction.  This means that the flows remain compatible around their bottleneck $\m{B}$.

We can now make a more general argument, applying to the entire remainder of $a$ and $b$ \strands\ of the flow. Consider any bulk point defined by the intersection of the $a$-foliation and $b$-foliation leafs (which, except for points lying on $\m{B}$, is uniquely specified by $\lambda_a$ and $\lambda_b)$. Inside $B$'s entangling region, $\ep^L_b(\lambda_b) < \ep^L_a(\lambda_a)=p_\alpha$ and simultaneously, in order for the leafs to intersect, the corresponding intervals cannot be nested, so $\ep^R_b(\lambda_b) < \ep^R_a(\lambda_a)$.  Moreover, the leafs cannot have multiple intersections. This means that at the intersection point, the normal for the $a$ leaf must bend more towards the $A$ region than the normal to the $B$ leaf.

Identical analysis for the remaining places where the \strands\ touch on one of the minimal surfaces then shows that {\it all} \strands\ bend away from each other.  Namely, once they are compatible on the minimal surfaces (i.e.\ satisfy the bottleneck equations \req{e:areaconds}), then they are compatible in the rest of the bulk --- in other words they generate cooperative flows.

\subsubsection{Single \locfol\ for all \strands, patched flows}
\label{sss:SecondConstr}

The previous construction required not just one, but five complete overlapping foliations to argue for cooperation (in particular to verify that the endpoints of foliating geodesics were such as to channel the \strands\ away from each other, cf.\ \fig{f:examplefols}c).
Since we ended up using only parts of these foliations, it might seem a bit more economical to consider just a single set of curves within the bulk which would generate all the \strands\ in one go.  Of course such a set cannot truly be a foliation in the strict sense that for every bulk point $p$ there exists a unique leaf of the foliation which contains $p$, because the topology of the \strands\ would necessitate some junctions in the leaves.  However, what really matters is not that we cover the entire bulk, but rather that the leaves don't intersect.  Once we exclude the junctions, this weaker (uniqueness) requirement is certainly possible to satisfy.  We will call such partial foliations (which we could think of as foliations of the bulk with some points removed) {\it \locfol s}.

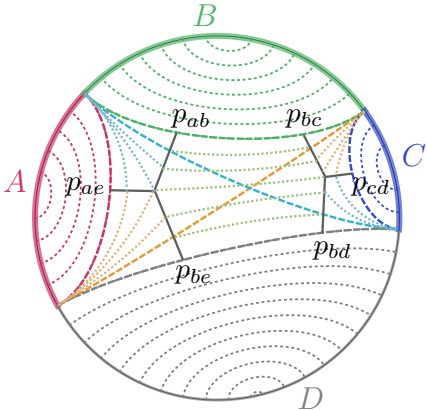
\begin{figure}[htbp]
\begin{center}
\includegraphics[width=2.1in]{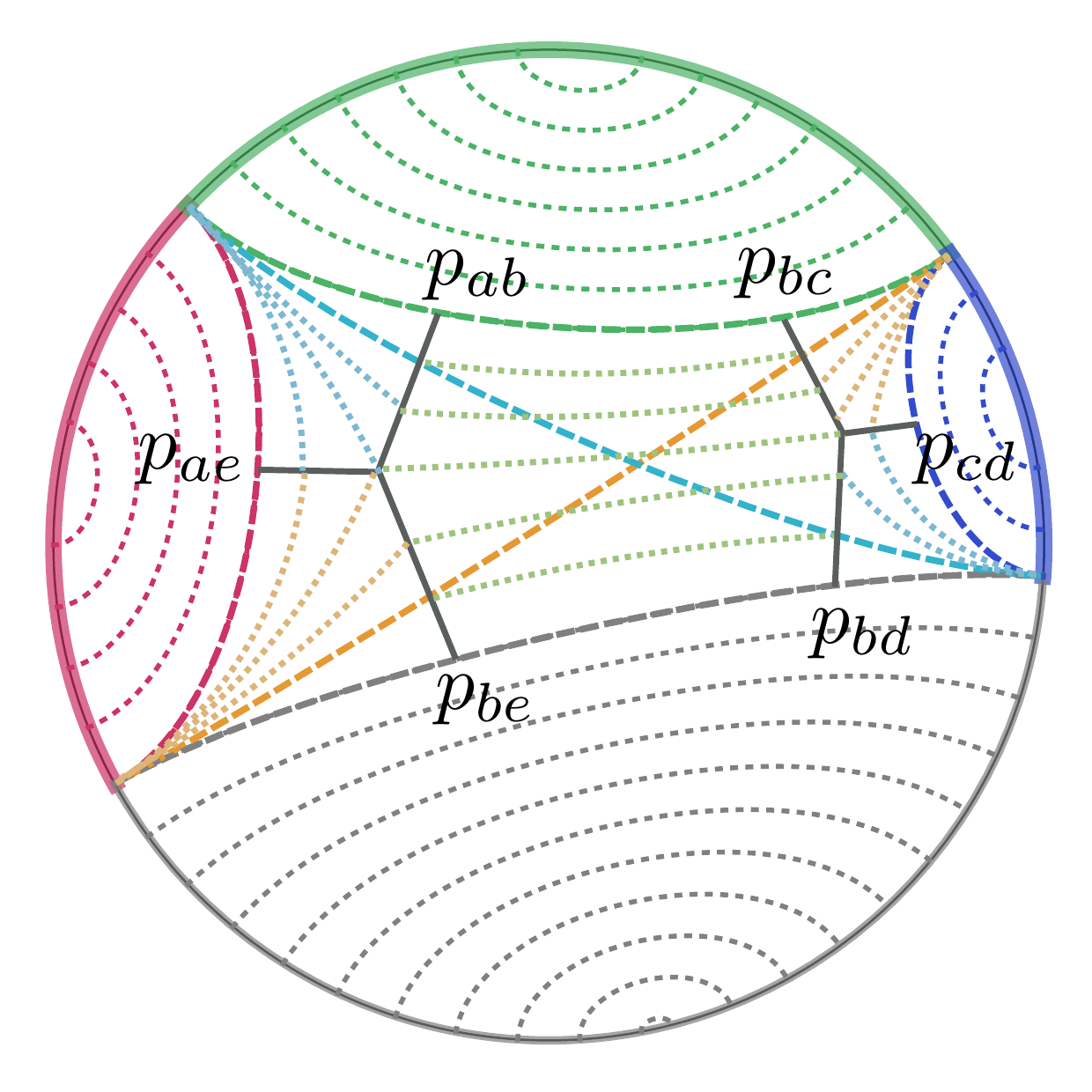}
\put(-162,80) {\makebox(20,20) {{\color{Acol}{$A$}}}}
\put(-90,143) {\makebox(20,20) {{\color{Bcol}{$B$}}}}
\put(-11,90) {\makebox(20,20) {{\color{Ccol}{$C$}}}}
\put(-50,-2) {\makebox(20,20) {{\color{Dcol}{$D$}}}}
\caption{A \locfol\ designed to obtain the \strands\ indicated in \fig{f:flowconstructs}b.  The 6 black line segments along with the 4 minimal surfaces break up the bulk into 9 different regions, each foliated by its own family of geodesics.}
\label{f:locfol}
\end{center}
\end{figure}

One (impractical) way to achieve a single \locfol\ of the bulk would be to take the set of foliations constructed in \sect{sss:FirstConstr}, cut them off at their respective \strand's edges (which are determined by the minimal surface partitions), and join the leafs through the remaining bulk gaps.  However, this offers no particular (either constructive or logical) simplification.  It also does not take the opportunity to avoid what is perhaps the most awkward feature of the previous construction, namely that the \strands\ between adjoining regions in  \sect{sss:FirstConstr} appear to start and end at the same boundary point.  Here we remedy this by a more convenient choice of \locfol, indicated in \fig{f:locfol}.

In the homology region\footnote{
By `homology region' for $X$, we mean the bulk region bounded by the given boundary region $X$ and its associated bulk minimal surface $\m{X}$, cf.\ footnote \ref{fn:EWdef} for a more covariant definition.
}
for each of the  boundary regions $A,B,C,D$, we use the foliation generated by symmetrically nested boundary intervals as in \fig{f:foliations}a.  All \strands\ $\v{X \fto Y}$ emanate from $X$'s midpoint, and automatically reach the corresponding minimal surface $\m{X}$ perpendicularly.  On the 4 minimal surfaces we pick a partition satisfying \req{e:areaconds}, which specifies 6 bulk points  labeled $p_{ab} \in \m{B}$, $p_{bc} \in \m{B}$, $p_{cd} \in \m{C}$, $p_{bd} \in \m{D}$, $p_{be} \in \m{D}$,  and $p_{ae} \in \m{A}$ as shown in   \fig{f:locfol} (the subscripts indicate where the given \strands\ separate, cf.\fig{f:flowregions}).
Consider the triangle formed by $p_{ab}$, $p_{ae}$, and $p_{be}$, and take its `center'  (to be specified in footnote \ref{fn:center}) point $p_{abe}$, and similarly for the $bcd$ junction.  Using these 6+2 points, construct 6 (geodesic) line segments joining the vertices of each triangle to its center point.  These segments partition the central quadrilateral region (bounded by the 4 minimal surfaces $\m{X}$) into 5 bulk regions, one for each \strand, thus partitioning the full Poincare disk into 9 regions.

To construct the five \strands, we proceed analogously to the previous construction, but with the crucial difference that we don't specify these geodesics by their endpoints on the boundary, but rather by their location along these interior line segments.  Concretely, the foliation for \strand\ $a$ is given by the family of geodesics between  $p_{AB} \equiv \partial A \cap \partial B$ and the interior bulk point $p_a^R(\lambda)$ lying in the segments $p_{ab}p_{abe}p_{be}$, with $\lambda = 0$ at $p_{ab}$ and $\lambda = 1$ at $p_{be}$.  Since all geodesics `fan out' from  $p_{AB}$, this is part of a foliation of the type indicated in \fig{f:foliations}b.  Moreover, our choice of $p_{abe}$ guarantees\footnote{
\label{fn:center}
The central point $p_{abe}$ is defined as any point which lies within the triangle formed by the threads passing through the three pairs of vertices in the construction of \sect{sss:FirstConstr}.  (Hence we retain a lot of freedom in this construction.)
Since these threads bend away from their tangents at the minimal surfaces $\m{X}$, they cannot intersect the line segments extending these tangents (pieces of which correspond to the segments $p_{ab}p_{abe}$ and $p_{abe}p_{be}$).
} that the corresponding threads from $\m{A \fto B}$ reach $\m{B \fto A}$ and vice-versa, without leaving this region.

The other three \strands\ between adjoining regions (namely $c$, $d$, and $e$), are constructed in identical fashion.  The remaining \strnd,  $b$, can actually be generated as a part of the foliation indicated in \fig{f:foliations}c. 
 And by the same argument as above, these threads joining  $\m{B \fto D}$ and $\m{D \fto B}$  cannot leave the hexagonal region $p_{ab}p_{abe}p_{be}p_{bd}p_{bcd}p_{bc}$.  This finishes the alternate construction of the \strands\ generating a \coop flow.

Notice that unlike the previous case discussed in \sect{sss:FirstConstr} which left 8 bulk regions unpopulated by any \strand, we now have only two unpopulated regions (the triangles mentioned above).\footnote{
The reader might at this point wonder why the latter appears to have manifestly smaller volume (in fact a finite one) than the former (which has infinite volume owing to some of the regions adjoining the boundary), despite the fact that in both cases we kept all our \strands\ maximally collimated.  This apparent discrepancy stems from the usual issue of the  UV divergences when dealing with the entire bulk spacetime. In particular, the threads of the present construction include those which `hug' the boundary and therefore utilize greater bulk volume than the threads of \sect{sss:FirstConstr}.
}
In fact, we can reduce this further to just a single unpopulated region, by using 
 only two intersecting line segments, $p_{ae}p_{cd}$ and say one joining midpoint of $\m{B \fto D}$ to midpoint of $\m{D \fto B}$.  The $a$,  $c$, $d$, and $e$ \strands\ would remain as before, but the $b$ \strand\ now gets split into two kinked sub-strands, one adjoining the $a$ and $e$ \strands\ and the other adjoining the $c$ and $d$ \strands.  The kinks appearing along the $p_{ae}p_{cd}$ segments pose no particular problem (the norm bound is satisfied everywhere) other than the flow field $\v{B \to D}$ being strictly-speaking undefined at the segment.  We can however locally smooth out the corners to avoid this.   
 
 The overall lesson is that there are many ways to construct cooperative flows generated by \locfol s (using minimal surface segments).  The resulting \strands\ are automatically disjoint and maximally collimated.   The gaps they leave behind in the bulk have no physical significance -- they (as well as precise position of the \strands\ themselves) may be viewed analogously to a gauge choice.  On the other hand, the freedom of satisfying the bottleneck equations (e.g.\ how far along the $\m{B}$ surface can the $b$ \strand\ shift) may have a more direct significance in terms of multipartite entanglement (cf.\ \cite{MattMMI}'s interpretation of $-\frac{1}{2}I_3$) which would be interesting to explore further.
  It is worth noting that these \strands\ constitute a special case of a {\it multiflow} discussed in \cite{MattMMI}, where the more general  $v_{ij}$'s  of \cite{MattMMI} in our construction all saturate the norm bound individually in some regions and vanish entirely in others.  We therefore provide an explicit construction of a special type of a `max multiflow' (in \cite{MattMMI}'s language).  
   
\section{Generalizations}
\label{s:generaliz}

In \sect{s:simple} we constructed \coop flows in the simplest class of examples: spatial slice of pure AdS$_3$ with boundary partitioning given by 4 adjoining intervals.  This starting point was chosen primarily for  ease of illustration, and   our arguments in fact apply much more broadly.  In this section we examine various natural generalizations.  
These entail considering more general partitions (still restricted to the form ${\cal H} = {\cal H}_A \otimes {\cal H}_B \otimes {\cal H}_C \otimes {\cal H}_D$ 
relevant for MMI), more general state (which modifies the asymptotically AdS bulk geometry and allows for time dependence), or even a different theory (including changing the number of dimensions and number of background spacetime components on which the field theory lives).

We conjecture that the basic construction outlined above can be adapted to any of the above-mentioned situations, describing a generic classical bulk spacetime satisfying physical energy conditions.  More specifically:
\begin{conjecture}
In any holographic CFT state whose gravitational dual corresponds to a classical bulk spacetime with entanglement entropy computed by the HRT prescription, a \coop flow for any boundary spatial partition exists and can be constructed by maximally collimated \strands\ confined to non-intersecting regions of a bulk Cauchy slice.
\end{conjecture}

Rather than attempting a full analysis (which we leave for future work), we provide some basic arguments and simple checks of our conjecture, intended more as a  plausibility argument than a proof.  Instead of considering the most general case from the outset, we address how our method can be adapted for each of these generalizations individually.  In particular, we consider generalizing the number of spatial regions in \sect{ss:partition}, generalizing the bulk geometry (within a class of time-reversal symmetric asymptotically AdS$_3$ spacetimes) in \sect{ss:geom}, generalizing to higher dimesions in \sect{ss:higherD}, and finally allowing for genuine time-dependence in \sect{ss:timedep}.

\subsection{Multiple sub-partitions}
\label{ss:partition}

Recall that the case of 4 adjoining intervals $A$, $B$, $C$, and $D$ discussed in the previous section has either $I(A:C)=0$ or $I(B:D)=0$, so the more immediate proof of \sect{ss:detour} suffices to guarantee the existence of \coop flows: in particular, we can pick $\v{A}$ and $\v{B}$ to be arbitrary and specify the remaining maximizer flows accordingly, without any need to split them into disjoint \strands.  We now demonstrate explicitly that in the more general situation with $I(A:C)>0$ and simultaneously $I(B:D)>0$, the `combed \strand' method of \sect{ss:constr} indeed applies.
We will first briefly outline a specific construction and then indicate what happens in full generality.

\begin{figure}[htbp]
\begin{center}
\includegraphics[width=2.1in]{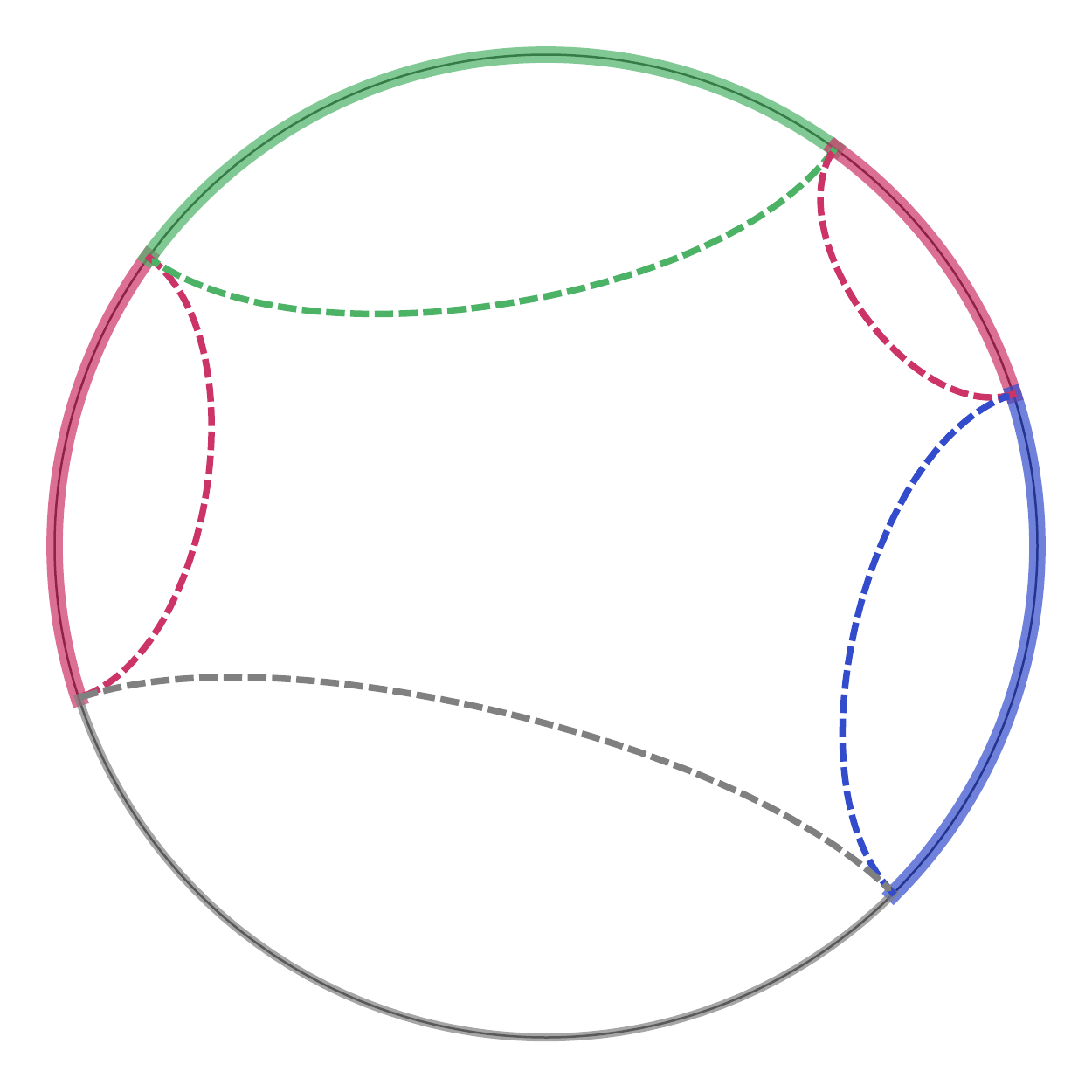}
\hspace{1.75cm}
\includegraphics[width=2.1in]{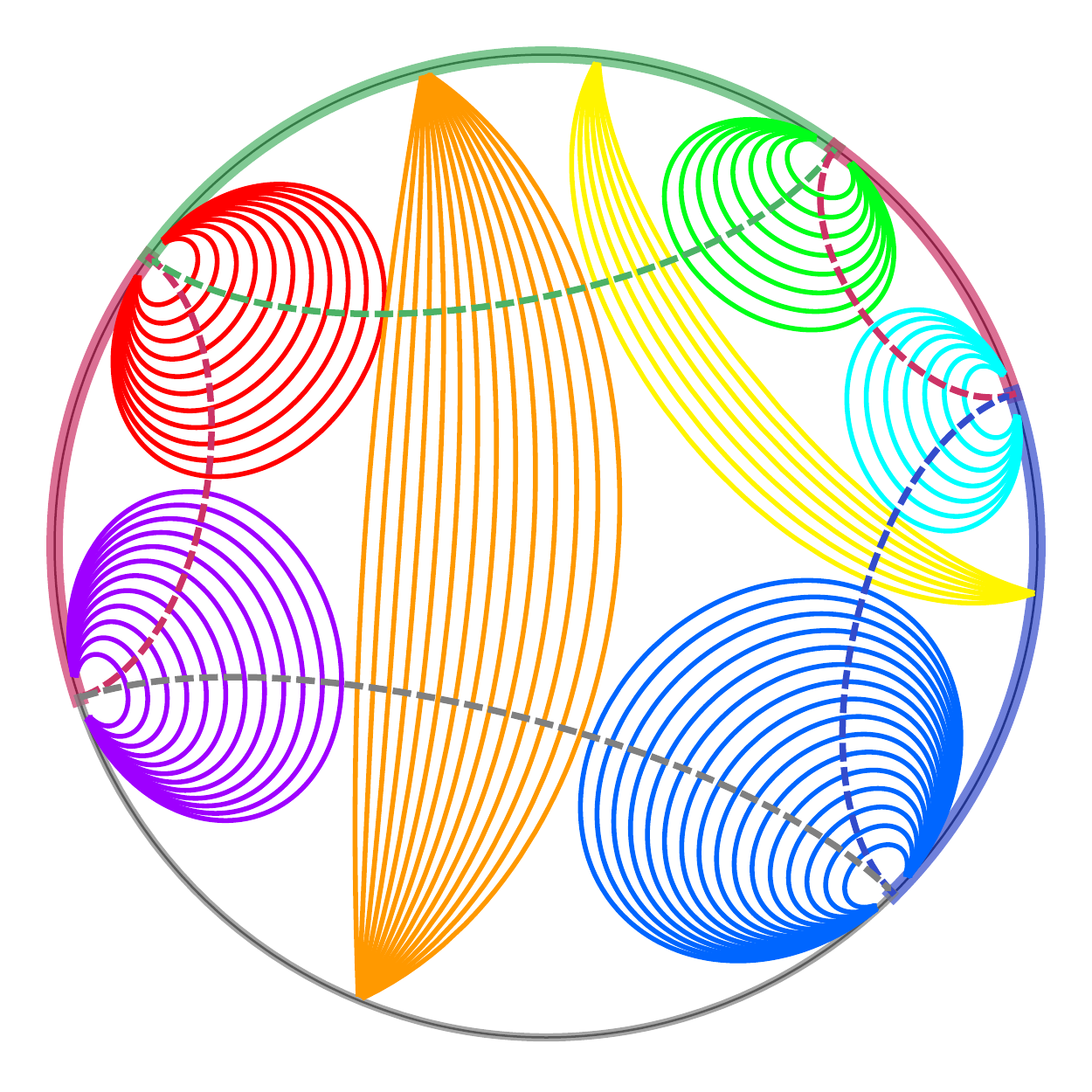}
\put(-163,80) {\makebox(20,20) {{\color{Acol}{$A_1$}}}}
\put(-20,110) {\makebox(20,20) {{\color{Acol}{$A_2$}}}}
\put(-90,143) {\makebox(20,20) {{\color{Bcol}{$B$}}}}
\put(-10,60) {\makebox(20,20) {{\color{Ccol}{$C$}}}}
\put(-50,0) {\makebox(20,20) {{\color{Dcol}{$D$}}}}
\put(-373,80) {\makebox(20,20) {{\color{Acol}{$A_1$}}}}
\put(-230,110) {\makebox(20,20) {{\color{Acol}{$A_2$}}}}
\put(-300,143) {\makebox(20,20) {{\color{Bcol}{$B$}}}}
\put(-220,60) {\makebox(20,20) {{\color{Ccol}{$C$}}}}
\put(-260,0) {\makebox(20,20) {{\color{Dcol}{$D$}}}}
\put(-125,90) {\makebox(20,20) {$a$}}
\put(-93,80) {\makebox(20,20) {$b$}}
\put(-65,90){\makebox(20,20) {$c$}}
\put(-53,105){\makebox(20,20) {$d$}}
\put(-38,83){\makebox(20,20) {$e$}}
\put(-60,46){\makebox(20,20) {$f$}}
\put(-125,55){\makebox(20,20) {$g$}}
\put(-330,74) {\makebox(20,20) {\footnotesize {\color{Acol}{$\m{A_1}$}}}}
\put(-305,105) {\makebox(20,20) {\footnotesize {\color{Bcol}{$\m{B}$}}}}
\put(-260,95) {\makebox(20,20) {\footnotesize {\color{Acol}{$\m{A_2}$}}}}
\put(-260,65) {\makebox(20,20) {\footnotesize {\color{Ccol}{$\m{C}$}}}}
\put(-295,50) {\makebox(20,20) {\footnotesize {\color{Dcol}{$\m{D}$}}}}
\put(-285,-10) {\makebox(0,0) {(a)}}
\put(-73,-10) {\makebox(0,0) {(b)}}
\vspace{0.4cm}
\caption{ {\bf (a)} An example of partitioning of the boundary space into 4 {\it subsystems} $A$, $B$, $C$, $D$, analogously to \fig{f:simple}, but now consisting of 5 simple {\it regions} as labeled.
{\bf (b)} Corresponding \strand\ construction analogous to \fig{f:flowconstructs}a, which delineates the associated bulk regions $a,b,c,d,e,f,g$ connecting the requisite boundary regions as specified in the text. }
\label{f:multAs}
\end{center}
\end{figure}
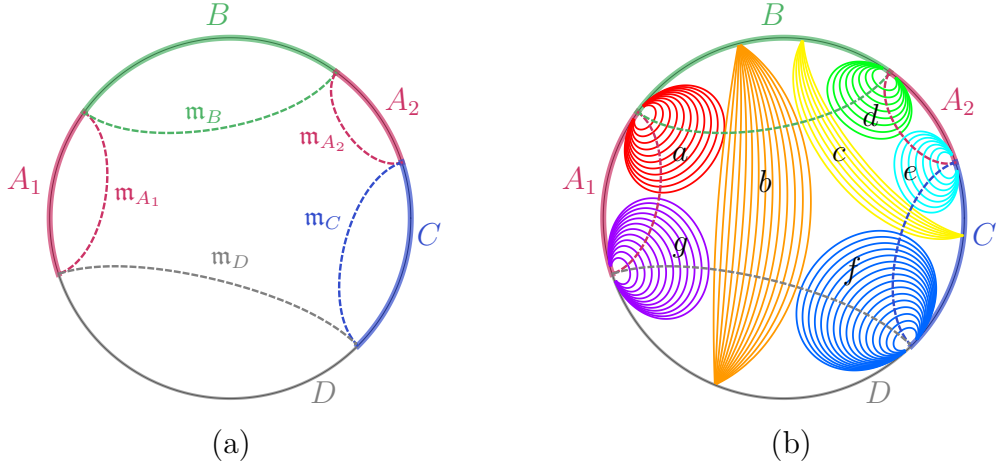

The simplest type of configuration which has all pairwise mutual informations non-vanishing is shown in \fig{f:multAs}a.    Here we take the $BD$ system to be larger than the $AC$ one, automatically implementing $I(B:D)>0$.  We also take $A=A_1 \cup A_2$ with $A_2$ adjoining $C$, so that $I(A:C)\ge I(A_2,C)=\infty$ (where the inequality follows by monotonicity of mutual information (SSA), and the RHS is the usual property of adjoining regions). This means that in decomposing the flows into \strands, we will necessarily have $\v{A_1 \fto B}$, $\v{A_1 \fto D}$, $\v{A_2 \fto B}$, $\v{A_2 \fto C}$,  $\v{C \fto D}$, as well as  $\v{B \fto D}$ in order to accommodate the non-vanishing mutual information between the corresponding intervals.  On the other hand, by the same argument as above, $I(A_1:A_2C)=0$ which (again by monotonicity along with non-negativity) implies that $I(A_1:A_2)=0$ and $I(A_1:C)=0$.  That in turn means that we do {\it not} need to turn on the $\v{A_1 \fto A_2}$ or $\v{A_1 \fto C}$ \strands. 

There are two additional \strands\ to consider:  $\v{A_2 \fto D}$ and $\v{B \fto C}$, but turning on both of these  maximally collimated \strands\ simultaneously  would not produce \coop flows, since the \strands\ necessarily intersect and therefore in the common region exceed the norm bound.
However, it is easy to see that we can always choose one of these \strands\ to vanish:  If $I(B:C)=0$, then we can set $\v{B \fto C}=0$ (so in this case \coop flows exist trivially).\footnote{
In particular, by argument of \sect{ss:detour}, we can set $\v{B}=\v{C}$ and $\v{A} = \v{ABC}$, which gives $\pv_\lc =   \v{A}$ and $\pv_\la =  \pv_\lb = \v{B}$, manifesting all as flows.
}
  On the other hand, if $I(B:C)>0$, then the complementary regions must be uncorrelated: $I(A_2:A_1D)=0$ which means that $I(A_2:D)=0$, so that we can set $\v{A_2 \fto D}=0$.  Let us then consider this latter case.
 
The first thing to check is that the requisite \strands\ can fit through the bottlenecks in a disjoint fashion. 
We have 5 minimal surfaces ($\m{A_1}$, $\m{B}$, $\m{A_2}$, $\m{C}$, and $\m{D}$) and 7 bulk regions $a,b,c,d,e,f,g$ containing the
7 \strands\ which form the building blocks for the flows, which we wish to prove to be mutually non-overlapping 
($a \supset \v{A_1 \fto B}$, 
 $b \supset \v{B \fto D}$, 
$c \supset \v{B \fto C}$,
$d \supset \v{B \fto A_2}$, 
$e \supset \v{A_2 \fto C}$,  
$f \supset \v{C \fto D}$, 
 and $g \supset \v{A_1 \fto D}$), cf.\ \fig{f:multAs}b (which shows the actual thread construction).
These \strands\ must partition the minimal surfaces so as to satisfy a set of bottleneck equations analogous to \req{e:areaconds}.  These consist of 5 equations (one for each interval $X$) of the form 
\begin{equation}
\area{\m{X}} = \sum_z \area{\mm{X}{z}} 
\label{eq:}
\end{equation}	
where $X$ denotes one of the 5 simple boundary intervals and $z$ denotes all the \strands\ with cross $\m{X}$, and 7 equations (one for each \strand\ z) of the form
\begin{equation}
\area{\mm{X}{z}} = \area{\mm{Y}{z}}
\label{eq:}
\end{equation}	
where $X$ and $Y$ is a pair of simple boundary intervals joined by a \strand\ $z$.  Since each \strand\ gives a partition for two minimal surfaces (associated with the regions the \strand\ connects), we have total of 14 partitions.  Hence we have 12 equations for 14 unknowns, which gives a 2-parameter family of solutions; we leave it as an easy exercise for the reader to write out the explicit equations and confirm that a solution must necessarily exist.

So far, we have ascertained that the 5 minimal surfaces can be partitioned so as to accommodate all the \strands\ correctly.  We can now employ the same arguments as in \sect{ss:constr} to verify that by using minimal surface foliations, the remainder of the \strands\ must bend away from each other.  Either of our explicit constructions is generalizable, but it's most convenient to adopt the first one, using 7 global foliations.  Since the reasoning presented at the end of \sect{sss:FirstConstr} was local in the sense of involving only two adjoining \strands, we can employ exactly the same argument in the present case.  
The explicit \strand\ construction is given in \fig{f:multAs}b.  As for the simple case, we can guide the \strnds\ using only the two foliation families depicted in \fig{f:foliations}b (for \strnds\ $a,d,e,f,g$) and \fig{f:foliations}c (for \strnds\ $b$ and $c$).  This choice again fixes the solution to the bottleneck equations uniquely.  We could also introduce more freedom by using only the separated endpoint form of the \strnds, or even more generally, a hybrid of the two methods.
 This proves the existence of a \coop flow for the 5-region case, where the methods of \sect{ss:detour} (utilizing vanishing mutual information) would have failed.

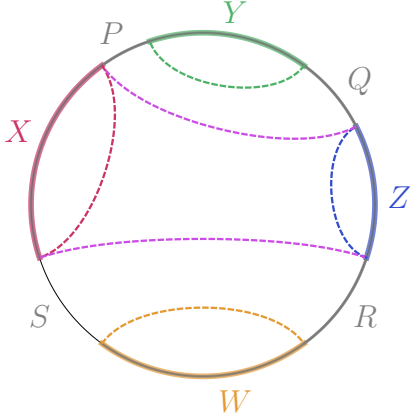
\begin{figure}[htbp]
\begin{center}
\includegraphics[width=2in]{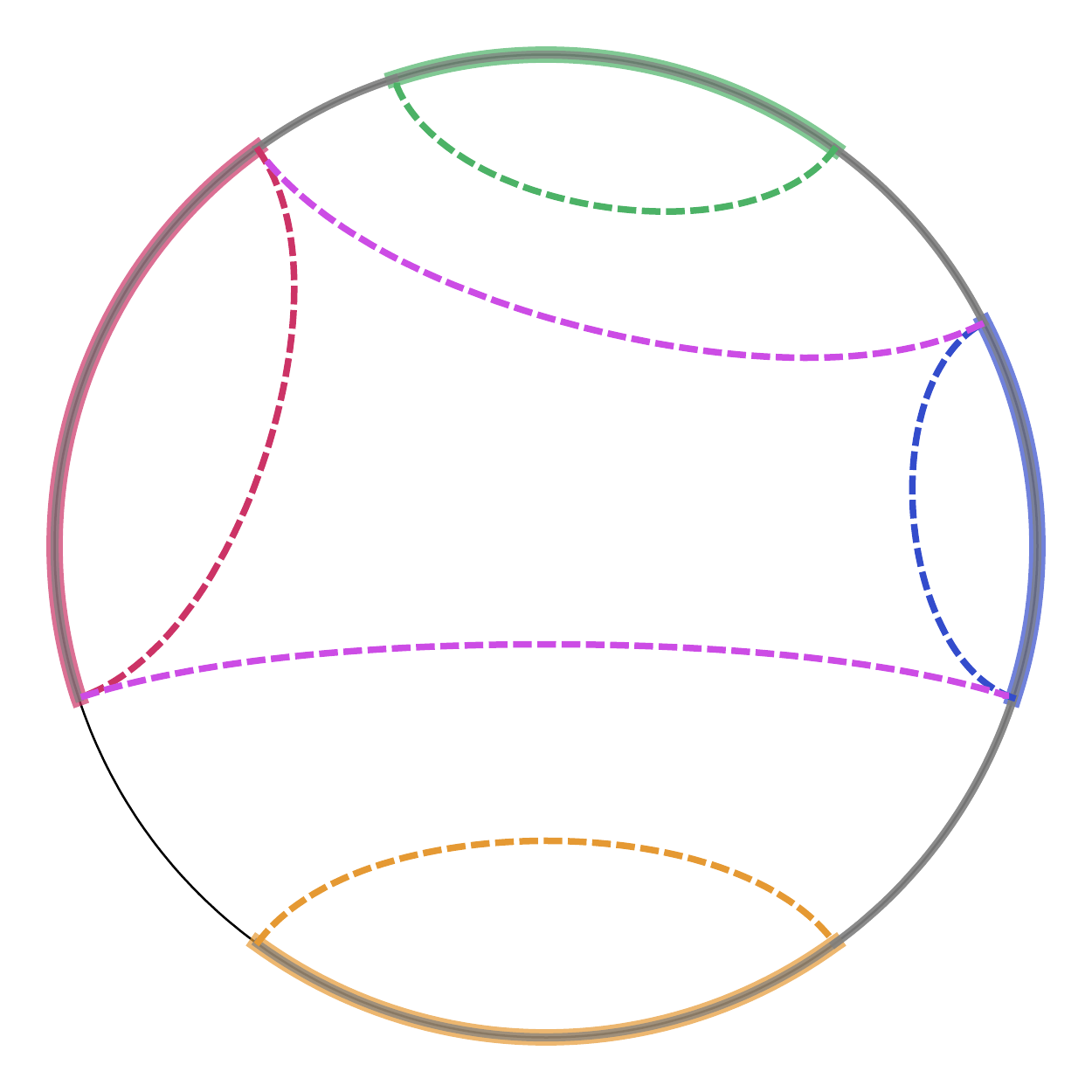}
\put(-152,90) {\makebox(20,20) {{\color{Acol}{$X$}}}}
\put(-70,135) {\makebox(20,20) {{\color{Bcol}{$Y$}}}}
\put(-8,65) {\makebox(20,20) {{\color{Ccol}{$Z$}}}}
\put(-70,-12) {\makebox(20,20) {{\color{ABcol}{$W$}}}}
\put(-117,127) {\makebox(20,20) {{\color{Dcol}{$P$}}}}
\put(-23,109) {\makebox(20,20) {{\color{Dcol}{$Q$}}}}
\put(-21,20) {\makebox(20,20) {{\color{Dcol}{$R$}}}}
\put(-144,20) {\makebox(20,20) {{\color{Dcol}{$S$}}}}
\caption{Generalization of \fig{f:simple} to allow our basic subsystems $A$, $B$, $C$ and $D$ to consists of arbitrarily many intervals, where we focus the potentially intersecting \strands\ for some 4 intervals $X$, $Y$, $Z$, and $W$ picked out of the basic set, separated by the remaining intervals, collected into $P,Q,R$, and $S$.}
\label{f:intersectstrands}
\end{center}
\end{figure}

It should be evident by now that we can generalize our construction to arbitrarily large set of boundary intervals $\{ A_i \}$, $\{ B_j \}$, $\{ C_k \}$, and $\{ D_\ell \}$, with arbitrary sizes and ordering.  To characterize the system of equations we have to solve, it is convenient to represent any such configuration by a graph, with nodes corresponding to the boundary regions and links to the \strands\ joining them.  In order for the \strands\ to be compatible, they cannot intersect, which means that the graph must be planar.\footnote{
While planarity is not a fundamental requirement in full generality (e.g., it can easily be violated in the higher dimensional case as discussed in \sect{ss:higherD}), here we use it to conveniently characterize the set of constraints.
}  To see that this is always possible, consider a pair of intersecting \strand\ candidates, say connecting  $X \leftrightarrow Z$ and  $Y \leftrightarrow W$ (in other words, the boundary contains the intervals $X,Y,Z$, and $W$ distributed somewhere along the circle in this order, but without any restriction on which subsystems they are part of), as indicated in \fig{f:intersectstrands}.  For labeling convenience, we'll collect the intervening regions into intervals labeled by $P,Q,R$, and $S$, though we don't require these to be non-vanishing.  

We can now generalize the previous argument to show that if $X$ is correlated with $Z$, then $Y$ cannot be correlated with $W$, and vice-versa:  Compare 
\begin{equation}
\area{\m{X}} + \area{\m{Z}}
\qquad {\rm versus} \qquad
\area{\m{PYQ}} + \area{\m{SWR}}
\label{eq:}
\end{equation}	
relevant for $S(XZ)$.
If the LHS is smaller than the RHS, then $I(X:Z)=0$.  On the other hand, if it's larger, then $I(PYQ:SWR) =0$ which by monotonicity and positivity implies that $I(Y:W)=0$  (and of course if the two sides are equal, both mutual informations vanish).  The consequence is that we never need to invoke the  $X \leftrightarrow Z$ and  $Y \leftrightarrow W$ \strands\ simultaneously.  This justifies our claim that the graph is planar.  

Let us now use the graph representation to count the number of equations and unknowns.  Suppose the graph has a total of $n$ nodes and $\ell$ links.  The number of unknowns $\mm{X}{z}$ corresponds to the number of partitions of the minimal surfaces, and since each \strand\ crosses two of the minimal surfaces, this is given by $2\ell$.  The number of equations is given by $n+\ell$, one for each minimal surface and one for each \strand.  This means that 
\begin{equation}
({\rm \# \ unknowns}) - ({\rm \# \ equations}) = \ell -n
\label{eq:}
\end{equation}	
The RHS is guaranteed to be non-negative: we must have at least as many links as nodes since all the adjoining regions have non-zero (in fact diverging) mutual information.  This already guarantees that we should find at least one solution.  Moreover, although nearly-equal size intervals could all have vanishing mutual information for any non-adjoining pair, more typically (if some regions are sufficiently large or sufficiently small), there will be additional correlations, making $\ell > n$, which would yield a $(\ell-n)$-parameter family of solutions:  the more regions are correlated, the more freedom we have in choosing the \strands.  In other words, we can shift threads between regions while maintaining the \coop nature of the flows.  
(On the other hand, we don't have arbitrary freedom, since planarity of the graph also puts an upper bound on the dimensionality of the space of solutions: $\ell \le 3n-6$.)

As before, once we have solved the minimal surface partitioning problem, the cooperativeness of the remainder of the \strands\ follows by the local arguments of \sect{sss:FirstConstr}.  Hence for any partitioning of the boundary, with arbitrarily many regions, of arbitrary sizes and ordering, we can always construct \coop flows.

\subsection{Exciting the geometry}
\label{ss:geom}

So far we have been working with a constant time slice of AdS$_3$ geometry, corresponding to the near-vacuum 
(or a slightly-excited state belonging to the same `code subspace' describing the same bulk geometry)  state of the CFT,  but in fact our arguments depended on this choice only in a very weak sense.   In particular, we only relied on $S(ABCD)=0$ and the minimal surface (i.e.\ geodesic) foliations of the bulk.  Clearly, as long as it remains true that to every boundary interval we can associate a unique bulk geodesic, that the set of these geodesics covers the entire spatial slice, and that they all end on a single boundary, our arguments will go through without modification.  Moreover, even in the more general spacetime where the above conditions fail, our arguments can still be suitably accommodated.

To see the unmodified part, note that minimal surfaces anchored on boundaries of nested regions cannot intersect.\footnote{
In more complicated situations considered below, we could have multiple components of a given minimal surface, say including a co-dimension-2  compact part, wrapping a horizon of a black hole.  In such cases, in the so-called plateaux regime \cite{Hubeny:2013gta}, these compact components associated with nested boundary regions, do coincide.  In higher dimensions, we could also contrive minimal surfaces to be tangent on some higher ($>2$)-codimensional locus.  However, neither of these is a problem for our construction.
}  
This follows by similar arguments as used in \sect{ss:setup}: if they did, then we could decrease their area further by smoothing out the corner at the intersection, thereby contradicting the assumption of minimality.
We will refer to this property as {\it nesting}.  Such a minimal-surface foliation then generates a maximally collimated flow, just as before.  Multiple foliations (or a \locfol) allows us to construct multiple \strands, whose compatibility is ensured by satisfying the requisite bottleneck equations (namely \req{e:areaconds} or their obvious generalization discussed in \sect{ss:partition}), since the argument for the non-intersection of the distinct \strands\ follows from nesting as before.

\begin{figure}[htbp]
\begin{center}
\includegraphics[width=2.1in]{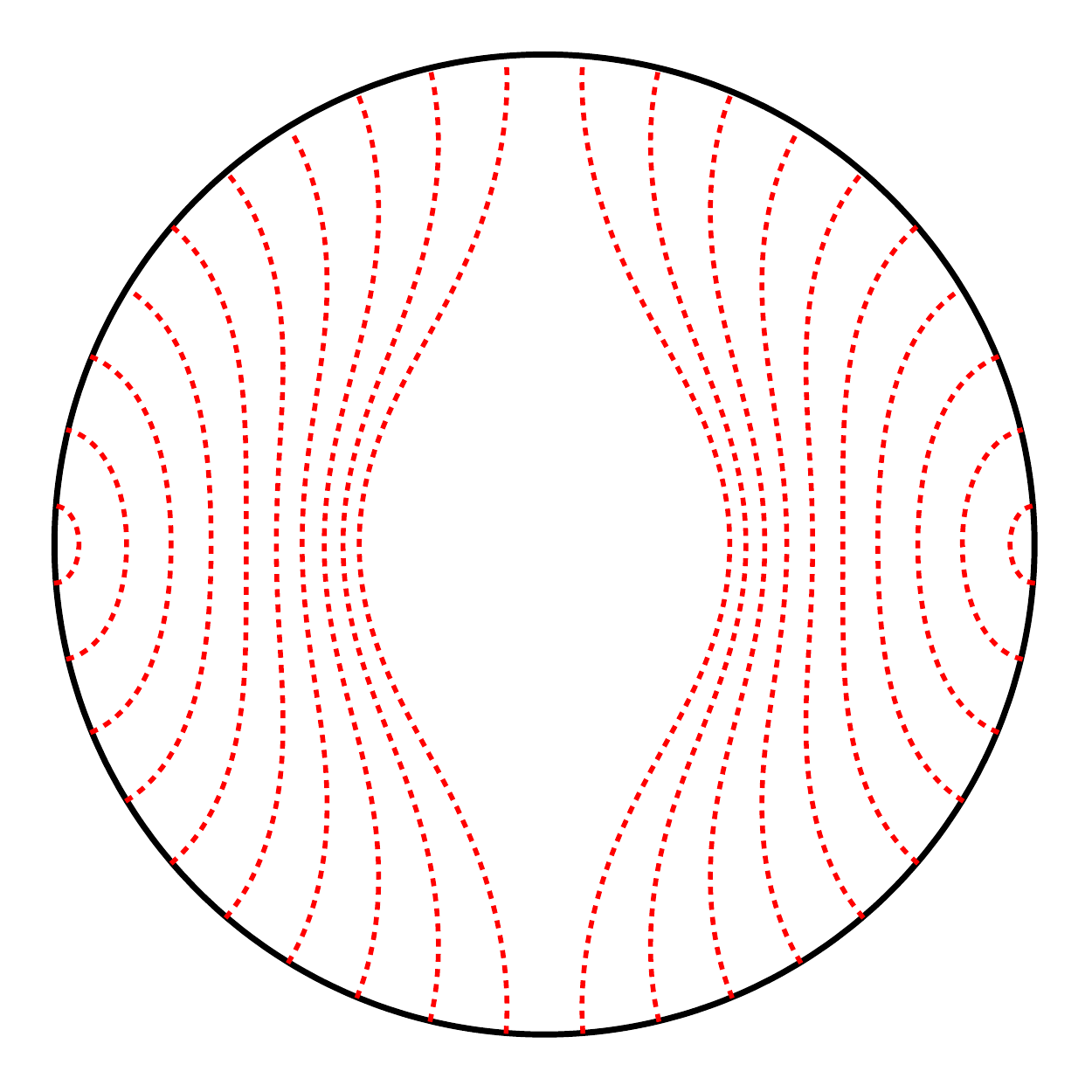}
\hspace{1.75cm}
\includegraphics[width=2.1in]{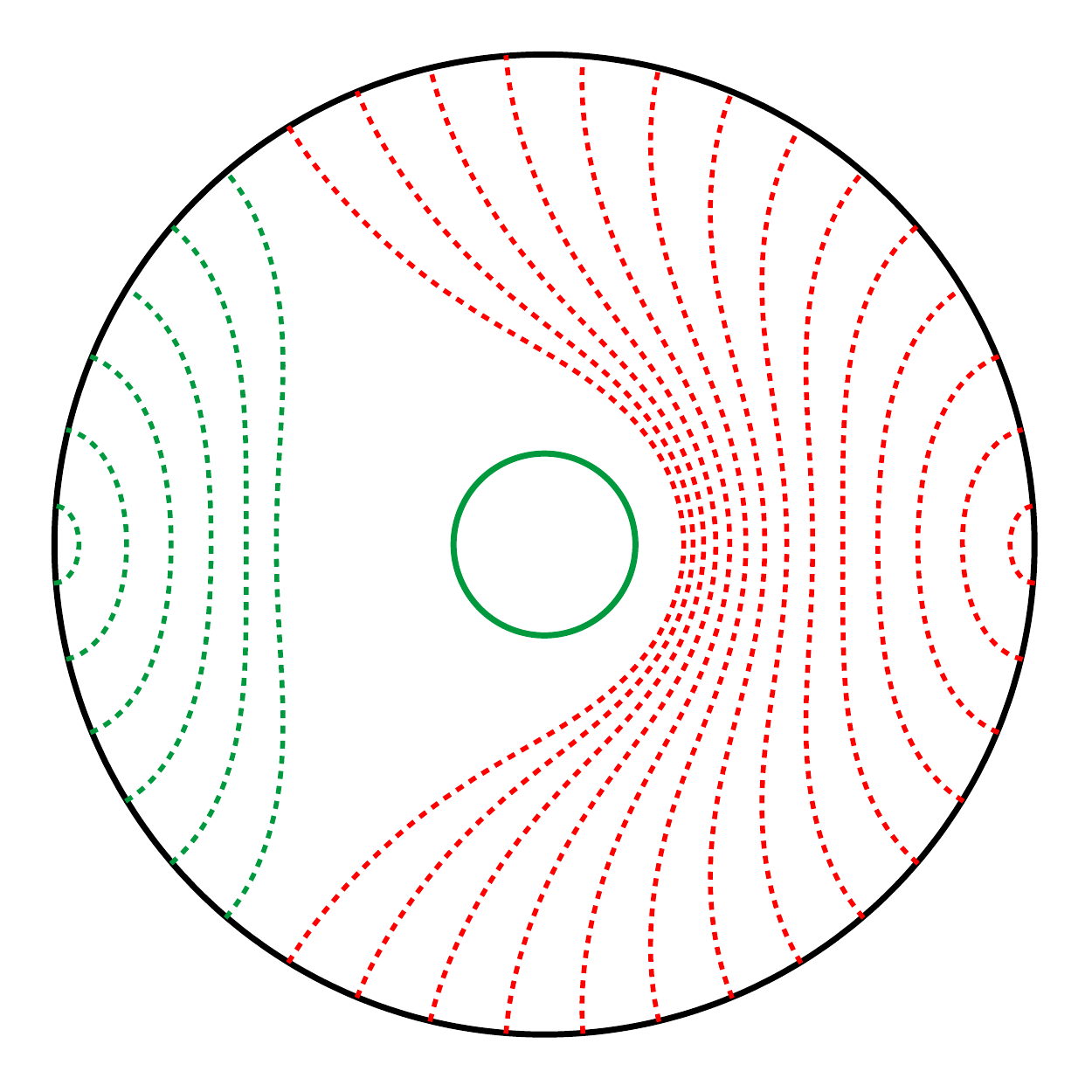}
\put(-285,-10) {\makebox(0,0) {(a)}}
\put(-73,-10) {\makebox(0,0) {(b)}}
\vspace{0.4cm}
\caption{Examples of bulk geometries wherein a boundary foliation does not generate a bulk foliation, in the sense that minimal surfaces leave behind `gaps'.  
{\bf (a)} A hypothetical topologically and causally trivial `compact star' geometry.  
{\bf (a)} A black hole (BTZ) geometry, with horizon indicated by the thick green curve in the center.  In both cases the dotted curves are globally minimal surfaces.  However it the black hole case, in order to satisfy the homology requirement, the minimal surfaces for a large enough region $(-\ph_0,\ph_0)$ consist of two disconnected components, the dotted green curve and the horizon.
}
\label{f:gap}
\end{center}
\end{figure}

Let us now consider sufficiently large deformations away from the AdS$_3$ geometry (but still keeping the metric time-reflection symmetric around the time of interest), such that minimal surfaces anchored on the boundary no longer foliate the bulk.  Although this can happen in a topologically and causally trivial way\footnote{
 For example for scalar solitons, see e.g.\ \cite{Nogueira:2013if,Gentle:2013fma}.
 (If the gap remains for all families of minimal surfaces, such as would be the case for spherically symmetric configurations of the above type, such regions would constitute what was dubbed `entanglement shadows' in \cite{Balasubramanian:2014sra}.  On the other hand, the temporal version of this effect is somewhat harder to achieve; under sufficiently strong assumptions, maximal-volume co-dimension one slices of the bulk, when anchored on boundary foliation, would provide a bulk foliation; cf.\ e.g.\  \cite{Couch:2018phr}.)
 }
 as sketched in  \fig{f:gap}a,
  the most interesting cases which have been well-studied involve black hole geometries.  As argued in e.g.\ \cite{Hubeny:2012ry}, minimal surfaces cannot penetrate the horizon of a static black hole, and minimality along with the homology constraint leads to a phase transition in the nature of the minimal surfaces.  In the so-called plateaux regime \cite{Hubeny:2013gta}, surfaces anchored on large enough region $X$ consist of two disconnected components, one corresponding to a minimal surface for the complementary region (on the same boundary) ${\overline X}$, and the other wrapping the black hole horizon bifurcation surface, as indicated in \fig{f:gap}b.  
 
First consider the simpler case, indicated in  \fig{f:gap}a.
This family of minimal surfaces guides a \strand\ between the left side $\ph = \pi$ and the right side $\ph = 0$.  Outside the central gap, this \strnd\ behaves as in the previous discussion.  Inside the gap, the \strnd\ actually has more freedom.  Imagine any spatial co-dimension one (in our case 1-d) `surface', anchored on the critical entangling surface at which the phase transition occurs (in  \fig{f:gap}a straddling the top and bottom, $\ph = \pm \pi/2$, e.g.\ a vertical line).  Any such surface has by construction larger area than the minimal surfaces at the edge of the gap, and therefore cannot impose a more stringent bottleneck for the flow.  We can foliate the gap by non-minimal surfaces, in which case the threads would simply spread out and refocus on the other side of the gap, or we could foliate parts of the gap by \locfol s which would keep the \strnd\ maximally collimated but possibly split into multiple strands.  Employing the latter method, we can recycle the constructions explained in \sect{sss:SecondConstr}.  Hence in this class of cases we can  obtain a \coop flow. 

Finally, in the topologically non-trivial case having horizons (cf.\ \fig{f:gap}b) and multiple boundaries, the main new ingredient is that not all the treads need to end on the same AdS boundary -- some can penetrate the horizon\footnote{
\label{fn:threadplateaux}
There is in fact a nice story regarding the bit thread version of plateaux:  There are three distinct regimes characterized by the size $2 \ph_0$ of the boundary region.  When the region is less than half the boundary $\ph_0 \le \pi/2$, the threads need not enter the black hole at all.  When it is larger than half but smaller than the critical size for the plateaux regime $\pi/2 < \ph_0 < \ph_{\rm crit}$, some threads must enter the black hole, but not necessarily saturate the norm bound there.  On the other hand, in the plateaux regime $\ph_0  \ge \ph_{\rm crit}$, the threads entering the horizon are necessarily maximally collimated.
}, 
and emerge on another boundary.  However, in such a case, we can redefine $D$ to include the entire spatial slice of the other boundary (as presaged in footnote \ref{fn:wormholes}), and proceed as in \sect{ss:partition} and the above arguments for not `minding the gap'.

The upshot is that we can still guide a \coop flow through any static $2+1$-dimensional asymptotically AdS bulk geometry, even when the minimal surface foliations leave behind gaps and the spacetime contains wormholes and boundaries (possibly ending up with more bulk regions which don't contain any \strnds).  To generalize further, we should consider what happens in higher dimensions, as well as what happens when we allow for general time dependence, which we address in the next two subsections.

\subsection{Higher dimensions}
\label{ss:higherD}

One crucial difference between the 2+1 dimensional spacetime we have hitherto considered and a higher-dimensional one is that while the former disallows simultaneous $AC$ and $BD$ flows on a constant-time slice (since the threads would then necessarily intersect), in the latter case both are allowed since they can go around each other in the extra direction.  Hence in higher dimensions we do not have to restrict to a subset of potential \strands, and relatedly, we have no natural ordering of the boundary regions.  If this were the only effect of higher dimensions, it would be clear that the feasibility of finding compatible flows necessarily improves with increasing dimension.  However, there is another consequence of higher dimensions, which could potentially have the opposite effect:  One might worry that the same bulk region has to be shared by many more boundary elements. 
We should therefore proceed more carefully.  

We can nevertheless utilize the same key feature as in the 3-dimensional case, namely that minimal surface foliations still comb the \strands\ in a maximally collimated fashion.  Recall that minimal surfaces satisfy the nesting property in any number of dimensions (indeed, we have been using the higher-dimensional language in order to make this generalization of our arguments more apparent).  There will now generally be gaps in the foliation as surfaces undergo phase transitions,\footnote{
In fact, minimal surfaces can have various non-trivial features (for example even for simply-connected boundary regions the bulk minimal surface can have handles) and distinct surfaces may be tangent along higher co-dimension subsets.
} however these can be dealt with in the same way as in \sect{ss:geom}.  Moreover, we can also more readily utilize the large volume near the boundary to lead most of the threads through this region in a more trivial fashion. 

More specifically, the bottleneck equations as such are not sensitive to dimensionality and their structure changes favorably (in the sense that each new \strand\ introduces 2 extra unknowns but only one extra equation).
Using nesting, we can then see that once the \strnds\ satisfy the compatibility conditions at the bottlenecks, the \locfol s can be chosen so as to guide them away from each other  in the rest of the spacetime.
These considerations suggest that it is indeed easier to construct \coop flows in higher spatial dimensions.  

One curious feature of changing spacetime dimensionality which affects threads is that while in two-dimensional space threads cannot swap endpoints (because that would require them to intersect), in three dimensional space they can swap endpoints in a topologically interesting way (i.e.\ they can be braided), and in higher dimensions the swapping is entirely trivial.  Since the RT prescription does not manifest this distinction, it would be interesting to see whether this has any actual physical implications.

\subsection{Time dependence}
\label{ss:timedep}

So far, we have been considering static\footnote{
One can generalize this class slightly to include bulk spacetimes with time reflection symmetry, provided we restrict our consideration to the specific time slice admitting this symmetry.
} configurations  which allowed us to compute the entanglement entropy via the Ryu-Takayanagi prescription   \cite{Ryu:2006bv} in terms of minimal surfaces $\m{X}$, or equivalently the Freedman-Headrick prescription \cite{Freedman:2016zud} in terms of bit threads $\v{X}$.  Even within the class of holographic theories with geometrical bulk dual, this is a severe limitation.  Not only are genuinely time-dependent configurations physically relevant and interesting in their own right, but in the context of holography they might also supply a key insight to structure of the holographic mapping and thereby elucidate the emergence of bulk spacetime.   This motivated generalizing of RT and FH prescriptions.

In general time-dependent situations, one can covariantize these prescriptions, which gives rise to the Hubeny-Rangamani-Takayanagi (HRT) prescription \cite{Hubeny:2007xt} in terms of extremal surfaces $\extr{X}$ which can also be recast into the Wall prescription \cite{Wall:2012uf} in terms of `maximin' surfaces,\footnote{
\label{fn:maximin}
More specifically, the Maximin prescription of \cite{Wall:2012uf} retains some useful features of RT, while restoring covariance in two steps:  First, on any bulk Cauchy slice $\Sigma$ whose boundary contains the entangling surface, we consider the area of a globally minimal surface homologous to the boundary region.  Second the entanglement entropy is given by maximizing this area over all Cauchy surfaces.  The resulting {\it maximin} surface realizing this maximum (under certain physically reasonable assumptions) is the HRT extremal surface, and any Cauchy slice within which this extremal surface minimizes the area is called a maximin slice.
} 
(allowing MMI to be proved in the time-dependent set-up), or the newer prescription \cite{covariantflows} in terms of  `covariant bit threads'.   The latter prescription builds on the methods of \cite{Headrick:2017ucz},  using convex relaxation on the maximin prescription and Lagrangian duality to express the entropy in terms of a flux of a divergence-free flow field $v$, whose norm bound gets replaced by a more global norm bound, integrated over temporal extent.  In other words, we still have 1-dimensional bit threads, which can however have a non-vanishing time component.  Nevertheless, one can argue that all threads must pass through the requisite extremal surface, in a maximally-packed manner (thereby reproducing the HRT prescription).  In fact, although the threads may be timelike-separated from each other elsewhere, it is likewise possible to flatten the full congruence to a single spatial surface (thereby reproducing maximin), thus allowing us to recycle the methodology developed for the static case. 

Let us now apply the covariant bit thread prescription to the situation at hand.  As a first step, we wish to consider a requisite family of extremal surfaces which would serve as `thread-guide' to a given \strand.
Since any normal congruence to extremal surfaces (whether spacelike or timelike or null) has by definition vanishing expansion, 
once we find a family of extremal surfaces which foliate a given Cauchy slice, we can guide our \strand\ along that slice in precisely the same manner as in the previous sections.   This is indeed possible thanks to the covariant version of nesting, which in the bulk gets uplifted to a property pertaining to full co-dimension-zero regions, called  entanglement wedges.\footnote{
\label{fn:EWdef}
Entanglement wedge (first introduced in \cite{Headrick:2014cta}) for a bulk region $X$ is  
the  domain of dependence of the bulk homology region $r_X$ (part of a bulk Cauchy slice such that $\partial \,r_{_X} = X - \m{X}$), or equivalently the set of spacelike-separated bulk points between $X$ and $\m{X}$.
}
In particular, entanglement wedges 
for nested boundary regions are themselves nested \cite{Wall:2012uf,Headrick:2014cta}, so the corresponding extremal surfaces themselves must lie on a common maximin slice.  A foliating family of boundary regions then generates a Cauchy slice locally foliated by the corresponding extremal surfaces.\footnote{
More generally the extremal surface family could leave gaps along the slice, but we can apply the same reasoning as in \sect{ss:geom} to this case. 
}

This suggests that one should likewise be able to repeat the above constructions for arbitrary set of boundary regions $\{ X \}$ and \strands\ $\{ z \}$, by localizing the threads to specific spatial slices of the bulk.  In particular, consider spacelike (maximin) slices $\Sigma_z$, foliated by extremal surfaces (anchored on a nested set of boundary regions whose entangling surfaces in turn foliate the boundary), which comb the bit threads in a maximally collimated fashion.  While different boundary-foliating families of entangling surfaces will generally give rise to different bulk Cauchy slices (which can then be generically timelike separated from each other), by construction these slices must intersect at the common extremal surfaces $\extr{X}$.   This means that all the \strands\ constructed from such foliations will encounter the joint bottlenecks at the $\{\m{X}\}$. To check for the consistency of the \coop\ flow at the bottlenecks, we use the same equations (analogous to \req{e:areaconds}) for the partitions of the extremal surfaces by $\mm{X}{z}$, and hence by the same arguments a solution to this set of equations exists.  Moreover,  entanglement wedge nesting also guarantees that two adjoining \strands\  $z$ and $z'$ passing normally through $\m{X}$ will be `steered away' from each other by the respective foliations as long as the boundary regions behave as prescribed for the static case.

Thus we can apparently uplift the arguments and constructions of \coop flows from the static case to a generic time-dependent case, thereby vastly enlarging its regime of validity to states for which HRT applies.  However, it would be interesting to consider whether we can obtain more interesting behavior by not flattening the \strands\ to Cauchy slices.  We leave this problem for future exploration.
 
\section{Discussion}
\label{s:disc}

Since the geometric proof of SSA and MMI using the HRT prescription does not manifest the profound difference between these two inequalities while the bit thread formulation of \cite{Freedman:2016zud} seems to distinguish them more strikingly, we have focused on proving MMI \req{e:sumMMI} using  bit threads.  In particular, we demonstrated the existence of disjoint \strands\ which constitute the building blocks of \coop flows, which in turn guarantee that MMI holds.  Although we detailed the construction  explicitly for AdS$_3$ geometry, we conjectured that our method of obtaining \coop flows applies to any bulk geometry for which HRT applies, and indicated how each type of generalization should work.  If this conjecture holds, it proves MMI in the bit thread formulation in equal generality as achieved previously  \cite{Hayden:2011ag,Wall:2012uf} using the RT/HRT prescription.

However, the connection to the first part of the title has hitherto remained tenuous.  In this section we revisit the relation to bulk locality, and speculate on some emerging lessons and open problems.

\paragraph{Bulk locality:}
Our proof of MMI involved constructing  maximally collimated disjoint \strands. 
To this end,  we have utilized minimal surface foliations of bulk regions, since their normal congruences have vanishing expansions.  The fact that minimal surfaces anchored on nested regions cannot intersect has ensured that these \strands\ are well-defined everywhere. 
However, here this nesting property has been applied more locally than previously.  For example the bit thread proof of SSA \cite{Freedman:2016zud} used nesting in the sense that there exists a common maximizer flow for nested regions.  In \cite{Headrick:2017ucz}, this was shown to be equivalent to the statement that bulk `homology regions' $\{ r_X \} $ corresponding to nested boundary regions $\{ X \}$ are themselves nested (or equivalently the associated minimal surfaces $\{ \m{X} \}$ don't intersect).  In both formulations, however, the starting assumption of nesting pertained to the {\it boundary}.
As explained in \cite{Freedman:2016zud}, this form of nesting does not suffice to prove MMI.  In retrospect, the reason for this seems to lie in that it is too global.  We needed an additional ingredient, namely further details of the bulk geometry.  In this sense, the fact that MMI holds for CFT states with geometric dual, and not necessarily in greater generality, naturally stems from the essence of the geometry: classically, it is local.

We however emphasize that MMI as such does {\it not} imply bulk locality, as has already been appreciated in \cite{Hayden:2011ag}; in fact even for 4-qubit systems (which one might think are maximally removed from anything resembling holography), MMI is generically upheld \cite{Rangamani:2015qwa}.   On the other hand, in the holographic context, it is interesting to ask what effect would stringy and quantum corrections have on our construction.  
From arguments based on RT  \cite{Hayden:2011ag}, one expects that MMI holds whenever the bulk is classical, regardless of the actual field equations governing the geometry at leading order in $N^2$, but not once the quantum effects become important.\footnote{
Even the subleading $1/N$ corrections in terms of bulk entanglement between the homology regions  \cite{Faulkner:2013ana} a-priori need not uphold MMI.
}  These expectations are consistent with the bit thread picture.
As shown in 
\cite{Harper:2018sdd}, the effect of adding higher-curvature terms to the gravitational action manifests itself in corrections to the norm bound: the bit thread thickness effectively changes in response to the local geometry and thread orientation.  However, since minimization (in the static case) is involved in determining the bottlenecks, one still retains a type of nesting, so one should still be able to guide the \strands\ through the bulk in a cooperative fashion.  On the other hand, finite-$N$ effects modify the divergencelessness condition \cite{Freedman:2016zud}, so threads can disappear and reappear at various points in the spacetime.  This invalidates our construction based on unbroken threads entirely, though a-priori the sign of this effect is unclear: one would expect a larger set of possibilities for a quantum flow. 

\paragraph{Role of monogamy:}

A key feature of quantum entanglement is the monogamy property:  maximum entanglement between two parties cannot be shared by a third party. 
As mentioned in the Introduction, this property would guarantee the negativity  of tripartite information $I_3$ (and hence MMI) if the mutual information only consisted of quantum entanglement.  Let us see if we can turn this intuition around, to further explore the implications of MMI being upheld in holography.
In particular, we would like to ask if there is some sense in which the bulk can be pictured as arising from purely monogamous entanglement.\footnote{
Indeed, phrased at this abstract level, this speculation is not dissimilar from previous statements in the literature such as spacetime being built up from quantum entanglement \cite{VanRaamsdonk:2010pw} or ER=EPR \cite{Maldacena:2013xja}. 
(We observe in passing that this is a-priori rather different from the ideas of the type discussed in e.g.\ \cite{Czech:2015qta} involving kinematic space where bulk geodesics played a crucial role.  We emphasize that our \strands\ are generally {\it not} composed of geodesics, since by geodesic deviation equation the latter wouldn't remain equidistant.)
}

Note that the bit threads, having only two ends and no junctions, are rather evocative of monogamy:  their endpoints cannot be shared by a third party just as quantum entanglement cannot be shared.  Could the threads then `weave the fabric of spacetime' in some concrete sense?  This is in fact reminiscent of some of the expectations regarding tensor networks in holography \cite{Swingle:2009bg}, though one would want this to operate at a much more local level.  An immediate objection to this naive picture is that the flows depend on not just the state of the full system but also on the specified  partition, whereas the bulk geometry is independent of the latter.  An associated puzzle is that if we interpret the threads as implementing a purely bipartite (as opposed to multipartite) structure of entanglement, one would conclude that the tripartite information then necessarily vanishes, i.e.\ MMI would necessarily be saturated,  contradicting explicit computations.

Let us examine this issue a bit further.  We have seen that although maximizer flows for correlated (disjoint) regions are incompatible, the \strands\ which these are built out of can nevertheless cooperate.  If we could play the same trick not only with  $\v{A}$, $\v{B}$, $\v{C}$, and $\v{ABC}$ as in our construction, but also simultaneously with $\v{AB}$, $\v{BC}$, and $\v{AC}$, then all the \strnds\ appearing in MMI  \req{e:sumMMI} would explicitly cancel.  We however manifestly cannot decompose flows into \strands\ in such a way as to reassemble all of these maximizer flows from the same set of \strnds, simply because the associated minimal surfaces, which the \strnds\ would have to traverse perpendicularly and with unit norm, intersect each other.  Interestingly, using the method of \sect{sss:SecondConstr} we could construct a \coop flow out of \strands\ which pass through each of the evoked extremal surfaces, including both $\m{AB}$ and $\m{BC}$, perpendicularly.\footnote{
For example, if we want to construct \coop flow which passes $\m{AB}$ perpendicularly, we merely need to extend the orange fanned-out foliations in \fig{f:locfol} into the central region, so that they both contain $\m{AB}$.  More specifically, we extend the foliation in the $(\partial A \cap \partial D) - p_{be} - p_{abe} - p_{ae}$ quadrilateral region into the lower portion of the central hexagonal region (below  $\m{AB}$), and similarly extend the $(\partial B \cap \partial C) - p_{bc} - p_{bcd} - p_{cd}$ quadrilateral region into the upper portion of the central hexagonal region (above  $\m{AB}$).  By a similar trick (with additional foliations fanning out from the central point), one could even guide our \strands\ perpendicularly through both $\m{AB}$ and $\m{BC}$, effectively splitting the $\v{B\fto D}$ \strnd\ to bypass the disallowed neighborhood of their intersection.}
However, this does not suffice for such \strnds\ to be maximizer flows for all of the associated regions, because the above construction would leave part of $\m{AB}$ and $\m{BC}$ (at the least in the vicinity of their  intersection) untraversed.  This explains how bit threads can indeed implement monogamy naturally while at the same time being compatible with non-trivial multipartite entanglement structure. 

\paragraph{Other entanglement relations and constructs:}
The above observations suggest that bulk locality should suffice to prove entanglement relations which are guaranteed to hold by monogamy (in the sense of mutual information being dominated by quantum entanglement as opposed to classical correlations).  One such interesting relation is the 5-party cyclic inequality (which can be thought of as a generalization of MMI)  \cite{Bao:2015bfa}.  Since this inequality has hitherto likewise eluded proof using bit threads, in Appendix \ref{s:cyclic} we offer a few comments on generalizing our methods to that case.  The argument of the inequality being guaranteed by existence of \coop flows is in fact directly analogous to that of \sect{s:proof}.  However, the actual construction of requisite \coop flows has to be modified from that discussed above, due to the novel feature of evoking simultaneous maximizer flows for overlapping regions, which (by essentially the same arguments as in the previous paragraph) cannot be achieved with a single set of disjoints \strnds.  Though it appears possible to achieve with multiple sets of strands, we leave the explicit construction for future work.
   It will be interesting to elucidate the additional property of bulk locality (if sufficient) to manifest cooperative flows in this case.  
   
 One might in fact hope that the full set of such higher-partite entanglement relations (defining the holographic entropy cone  \cite{Bao:2015bfa}) would contain sufficient set of specifications of the requisite entanglement structure to identify geometric states directly, which partly motivates the search for further such relations.\footnote{
These are known exhaustively for only up to 5 parties, and generalizations of some of the known relations (such as the 5-party cyclic inequality) to arbitrarily large (but in the cyclic case odd) number of parties already guarantee infinitely many independent inequalities \cite{Bao:2015bfa}.  
(The upcoming work initiated by \cite{Entropycone} develops a program to generate further such relations and corresponding information quantities efficiently.)
 }  It would be interesting to see if, for all relations following from the existence of \coop flows, the requirement of realizability of these via maximally collimated disjoint \strands, in fact aids in this quest.
 
Let us close with a more general comment.  Originally the bit thread prescription of \cite{Freedman:2016zud} was formulated with the explicit goal to provide an alternate prescription for entanglement entropy.  In the present work we have utilized bit threads accordingly, since we have only considered quantum information theoretic constructs built out of entanglement entropy (most generally these comprise the information quantities mentioned above in the context of the holographic entropy cone). 
However, there are many other interesting and useful information theoretic quantities which are {\it not} directly constructed from entanglement entropy.  
As such, a-priori, the bit threads need not have any useful relation to such quantities.
However, we find the above constructions of \strands\ highly encouraging for finding wider applications of bit threads.  In particular, our constructions are rather reminiscent of the holographic entanglement of purification proposed in \cite{Takayanagi:2017knl}.  Furthermore, since bit threads might also be cognizant of certain aspects of the quantum / classical separation, it would be interesting to explore whether they bear any connection to concepts such as relative entropy of entanglement, quantum discord, robustness, entanglement of distillation, etc..  These in turn might reveal useful insights into the emergence of bulk spacetime in holography.

\section*{Acknowledgements}

It is a great pleasure to thank
Matt Headrick,
Juan Maldacena,
Mukund Rangamani, 
Max Rota,
and Tadashi Takayanagi
for stimulating discussions.
I would also like to thank Matt Headrick for comments on an earlier version of the draft.
I am grateful to 
 the Kavli Institute for Theoretical Physics in Santa Barbara, 
 the Centro Atomico Bariloche, 
 ICTS-TIFR in Bengaluru,
 the Galileo Galilei Institute in Florence, 
 and the Yukawa Institute for Theoretical Physics at Kyoto University,
 for hospitality during various stages of this project. 	
This work was supported by  U.S. Department of Energy grant DE-SC0009999 and by funds from the University of California. 

\appendix
\section{5-party cyclic inequality}
\label{s:cyclic}

Hitherto we have focused attention on MMI, which is an entropy inequality that involves partitioning the Hilbert space into 4 parts, which we denoted $A$, $B$, $C$, and $D \equiv \overline{ABC}$.  However, our geometric constructs as such didn't rely on this number of partitions, so one would expect the methodology to be applicable more universally.  As an illustrative example, consider the 5-region cyclic inequality, which can be thought of as a generalization of MMI \cite{Bao:2015bfa}.  For 6 partitions, $A$, $B$, $C$, $D$, $E$, and the purifier $F \equiv \overline{ABCDE}$, the cyclic inequality states
\begin{equation}
\begin{split}
S(ABC) + & S(BCD) + S(CDE) +S(DEA) + S(EAB) \ge \\ & S(AB) + S(BC) + S(CD) + S(DE) + S(EA) + S(ABCDE)
\end{split}
\label{e:cyclic}
\end{equation}	
Performing a similar exercise as in \req{e:LHSbound}, we can bound the LHS of \req{e:cyclic} by
\begin{equation}
\begin{split}
& \int_{ABC} \v{ABC} + \int_{BCD} \v{BCD} + \int_{CDE} \v{CDE} + \int_{DEA} \v{DEA} + \int_{EAB} \v{EAB} \\
&\ge
\int_{ABC} v_1 + \int_{BCD} v_2 + \int_{CDE} v_3 + \int_{DEA} v_4 + \int_{EAB} v_5 \\ 
&= 
\mbox{\footnotesize{$
\int_{A} (v_1 + v_4 + v_5) + 
\int_{B} (v_1 + v_2 + v_5) + 
\int_{C} (v_1 + v_2 + v_3) + 
\int_{D} (v_2 + v_3 + v_4) + 
\int_{E} (v_ 3+ v_4 + v_5) $}}
\end{split}
\label{e:LHSboundcyclic}
\end{equation}
where the inequality holds for any flows $v_1, v_2, v_3, v_4, v_5$.
We want to make the last line of \req{e:LHSboundcyclic} to be equal to the RHS of \req{e:cyclic}, i.e.
\begin{equation}
\begin{split}
& \int_{AB} \v{AB} + \int_{BC} \v{BC} + \int_{CD} \v{CD} + \int_{DE} \v{DE} + \int_{EA} \v{EA} + \int_{ABCDE} \v{ABCDE} =  \\
&=
\int_{A}  ( \v{AB} + \v{EA} + \v{ABCDE}) +
\int_{B}  ( \v{AB} + \v{BC} + \v{ABCDE}) + \\
&+
\int_{C}  ( \v{BC} + \v{CD} + \v{ABCDE}) +
\int_{D}  ( \v{CD} + \v{DE} + \v{ABCDE}) + \\
&+
\int_{E}  ( \v{DE} + \v{EA} + \v{ABCDE})
\end{split}
\label{e:RHSboundcyclic}
\end{equation}
which we can achieve if we can define
\begin{equation}
\begin{split}
v_1 &= \frac{1}{3} \left( \v{AB}  +  \v{BC} +  \v{CD} -2 \v{DE}  +  \v{EA} +  \v{ABCDE} \right) \\
v_2 &= \frac{1}{3} \left( \v{AB}  +  \v{BC} +  \v{CD} +  \v{DE}  -2 \v{EA} +  \v{ABCDE} \right) \\
v_3 &= \frac{1}{3} \left( -2\v{AB}  +  \v{BC} +  \v{CD} +  \v{DE}  +  \v{EA} +  \v{ABCDE} \right) \\
v_4 &= \frac{1}{3} \left( \v{AB}  -2 \v{BC} +  \v{CD} +  \v{DE}  +  \v{EA} +  \v{ABCDE} \right) \\
v_5 &= \frac{1}{3} \left( \v{AB}  +  \v{BC} -2  \v{CD} +  \v{DE}  +  \v{EA} +  \v{ABCDE} \right)
\end{split}
\label{e:cyclicvidef}
\end{equation}
such that all the $v_i$'s on the LHS of \req{e:cyclicvidef} are flows.  

\begin{figure}[htbp]
\begin{center}
\includegraphics[width=2.5in]{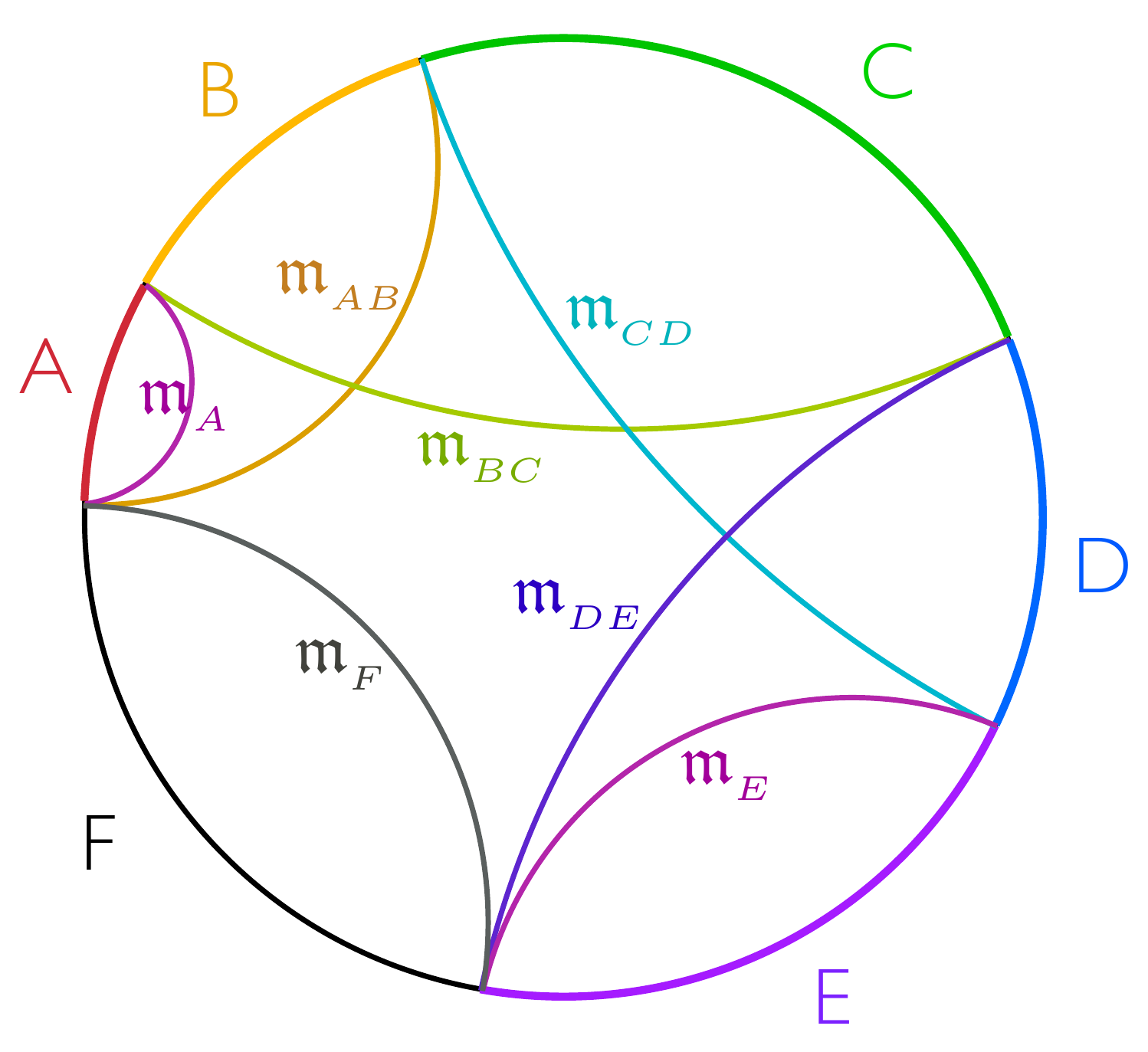}
\caption{Analogous to \fig{f:simple}: An example of partitioning of the boundary space into six regions $A$, $B$, $C$, $D$, $E$, and the complement $F \equiv \overline{ABCDE}$, as labeled.  The corresponding minimal surfaces for $AB$, $BC$, $CD$, $DE$, and $EA$  are also indicated.  The $AC$ minimal surface assumes \req{e:BCDFneck}. }
\label{f:cyclicminsurf}
\end{center}
\end{figure}

Let us consider a simple example of this scenario, analogous to \sect{s:simple}, with $A$, $B$, $C$, $D$, and $E$ being adjoining intervals on AdS$_3$ boundary, as indicated in \fig{f:cyclicminsurf}, which also shows the minimal surfaces for $AB$, $BC$, $CD$, $DE$, and $EA$.
Note that to draw the relevant minimal surface for $AE$ as $\m{A} \cup \m{E}$, we assumed that 
\begin{equation}
\area{\m{BCD}}+\area{\m{F}}  \ge \area{\m{A}}+\area{\m{E}} \ .
\label{e:BCDFneck}
\end{equation}	

We immediately observe a novel aspect compared to the MMI case:  Unlike the previous situation (cf.\  \fig{f:flowregions}) where we had a single set of \strands\ which we could use for all the $v_i$'s, we now have expressions involving flows through intersecting minimal surfaces.  This means that a single set of \strands\ does not suffice, i.e.\ we cannot have a fixed set of bit threads which would maximize the flow through $AB$ and through $BC$ simultaneously since $\m{AB}$ intersects $\m{BC}$ with different normal direction.

Nevertheless, if we evaluate the five expressions \req{e:cyclicvidef} on the six bottlenecks, they appear to be mutually compatible.   This suggests that a  \coop flow construction should be possible, albeit utilizing a more complicated method, evoking multiple superimposed \strands.  We however we leave  an explicit construction for future work.


\providecommand{\href}[2]{#2}\begingroup\raggedright\endgroup

\end{document}